\documentclass[preprint,showpacs,preprintnumbers,amsmath,amssymb,prl,longbibliography]{revtex4-1}

\usepackage{graphicx}
\usepackage{dcolumn}
\usepackage{bm}
\usepackage{color}
\usepackage{siunitx}
\usepackage{multirow}
\usepackage{makecell}
\usepackage[caption=false]{subfig}
\usepackage{hyperref}

\hypersetup{
    bookmarks=true,         
    unicode=false,          
    pdftoolbar=true,        
    pdfmenubar=true,        
    pdffitwindow=false,     
    pdfstartview={FitH},    
    pdftitle={My title},    
    pdfauthor={Author},     
    pdfsubject={Subject},   
    pdfcreator={Creator},   
    pdfproducer={Producer}, 
    pdfkeywords={keyword1, key2, key3}, 
    pdfnewwindow=true,      
    colorlinks=true,       
    linkcolor=blue,          
    citecolor=blue,        
    filecolor=blue,      
    urlcolor=blue           
}

\begin{document}

\title{Dynamical amplification of magnetoresistances and Hall currents up to the THz regime}
\author{Filipe S. M. Guimar\~aes$^{1}$}
\email{f.guimaraes@fz-juelich.de}
\author{Manuel dos Santos Dias$^{1}$} 
\author{Juba Bouaziz$^{1}$}
\author{Antonio T. Costa$^{2}$}
\author{Roberto B. Muniz$^{2}$}
\author{Samir Lounis$^{1}$}
\affiliation{$^1$ Peter Gr\"unberg Institut and Institute for Advanced Simulation, Forschungszentrum J\"ulich \& JARA, J\"ulich D-52425, Germany\\ $^2$ Instituto de F\'{\i}sica, Universidade Federal Fluminense, Niter\'oi, Brazil}

\date{\today}

\begin{abstract}

Spin-orbit-related effects offer a highly promising route for reading and writing information in magnetic units of future devices. 
These phenomena rely not only on the static magnetization orientation but also on its dynamics to achieve fast switchings that can reach the THz range. 
In this work, we consider Co/Pt and Fe/W bilayers to show that accounting for the phase difference between different processes is crucial to the correct description of the dynamical currents.
By tuning each system towards its ferromagnetic resonance, we reveal that dynamical spin Hall angles can non-trivially change sign and be boosted by over 500\%, reaching giant values.
We demonstrate that charge and spin pumping mechanisms can greatly magnify or dwindle the currents flowing through the system, influencing all kinds of magnetoresistive and Hall effects, thus impacting also dc and second harmonic experimental measurements.

\end{abstract}

\maketitle

\section*{Introduction}

The interrelation between magnetic properties and electric currents has been studied for more than a century \cite{{Thomson:1856jw},{Jan:1957gk}}. 
The electric resistance of magnetically ordered systems, for example, depends on the direction of their magnetization, a phenomenon given the general name of magnetoresistance. 
Systems exhibiting relatively large magnetoresistance are used in sensing devices for reading information stored in magnetic media.
Understanding how the spin of the electron can affect its propagation in materials with strong spin-orbit interaction is the main focus of various recent developments.
A plethora of novel magnetoresistive effects have been discovered in the last years: spin Hall \cite{Nakayama:2013gs,Chen:2013kf}, unidirectional spin Hall \cite{Avci:2015jp} and, independently, linear spin Hall \cite{Olejnik:2015bc}, as well as Hanle \cite{Velez:2016bm}, and Rashba-Edelstein  effects \cite{Nakayama:2016gp}. 
A common characteristic shared by all of them is the appearance of non-equilibrium spin polarizations accumulated at the interfaces/edges of those systems due to the flow of an electric current. 

This spin accumulation can also be used to excite spin waves in adjoining magnetic units, by means of the so-called spin-orbit torques, when time-dependent (ac) currents are applied \cite{Fang:2011ch,Garello:2013fa,Freimuth:2014kq,Freimuth:2015fq,Guimaraes:2015fl}. 
These torques can be effective even at room temperatures and may be used for magnetization switching purposes \cite{Miron:2011gd,Pai:2012ef,Liu:2012dga,Garello:2014bi,Ciccarelli:2016ju,Fukami:2016ca}. 
In certain metallic systems, the induced spin-orbit torques and magnetoresistive effects can reach fairly high intensities \cite{{MihaiMiron:gv},{Kim:2016di}}.
Although the connection between the magnetoresistance and static torques is well-established, a dynamical connection between these two effects is missing.
Unraveling how the induced dynamics influences the magnetoresistance and the Hall currents are essential questions, especially for prospective spintronics applications within the THz frequency range.

The effective magnetic fields that drive the magnetization into precession in those systems are obtained experimentally through second order signals, either dc (rectified) \cite{Fang:2011ch,{Liu:2012dga},{Kurebayashi:2014br}} or doubled-frequency (second-harmonic) voltages \cite{{Garello:2014bi},{Avci:2014fj},{Hayashi:2014bd}}. These measurements depend both on the transverse magnetization components, and on the magnetoresistance (or Hall resistance). 
The latter is assumed to be independent of both the frequency of the oscillatory electric field and intensity of the bias magnetic field that may be swept through the ferromagnetic resonance.
The measurements are usually described by simple functional forms \cite{Jan:1957gk,McGuire:1975jk}, and the characteristics of the magnetization dynamics are studied within a semi-classical approach.
In addition, even though the effect of the magnetization precession on second order signals is commonly accessed, measurements of its feedback in the first order signal are scarce \cite{Kupferschmidt:2006ho,Kovalev:2007hk}, specially for the in-plane geometry \cite{{Tserkovnyak:2014jd},{Ciccarelli:2014bl}}, where the pumped currents may give rise to important contributions yet to be explored. 

In this work, we investigate the intrinsic feedback of the current-induced magnetization dynamics to the electric current that drives it.
By applying a time-dependent electric field to ferromagnetic/heavy metal bilayer systems such as Co/Pt(001) and Fe/W(110), we demonstrate the emergence of an enhanced \textit{dynamical} magnetoresistance originating from spin and charge pumping. This can be one order of magnitude larger than the dc values.
The magnetization dynamics also induces currents along the transverse directions that substantially influence Hall effects. 
We show that charge to spin conversion in the dynamical regime hinges crucially on signal dephasing, implying that the spin Hall angle is a complex quantity. 
It is characterized by an amplitude and a phase difference, contrary to what is assumed in common phenomenological approaches.
This complex nature allows striking variations --- up to 500\% --- on the generated transverse spin currents.
Our approach is based on realistic descriptions of the electronic structure of the system. 
The theory describes collective spin excitations in the presence of spin-orbit interaction, and naturally takes into account all the intrinsic spin-orbit-related mechanisms, with no need for any adjustable parameters.

In the systems we have considered, we further show that magnetic excitations are the main source --- but not the only one --- of the dynamic contributions. 
They can be manipulated by varying the frequency and intensity of the applied fields, which can also affect second order signals measured in experiments.
Moreover, since the spin-pumping contribution is switched on or off by changing the magnetization direction, it leads to an angular dependence of the currents that deviates starkly from the static counterpart. 
This change in current signals provides an efficient way of manipulating magnetic states and establish a powerful ingredient for the development of spintronic devices.

\section*{Results}

\subsection*{Conceptual framework}

We consider metallic bilayer heterostructures consisting of a ferromagnetic metal deposited on a heavy metal substrate.  
The systems are subjected to a time-dependent uniform electric field given by $\mathbf{E}(t) = \operatorname{Re}[E_0e^{-i\omega t}\mathbf{\hat{x}}]$, where $E_0$ is the field amplitude, and $\omega$ its angular frequency.
An electric current may be measured along the longitudinal (x) or along the transverse (y) direction. The induced first order electric currents flowing through the system can be summarized in vector form as
\begin{equation}
\begin{split}
\mathbf{I}_\text{C}(t)  &= \operatorname{Re}\{(I_{\omega,\|}\mathbf{\hat{x}}+I_{\omega,\perp}\mathbf{\hat{y}})E_0e^{i\omega t}\}\ .
\end{split}
\end{equation}
where $I_{\omega} = I'_{\omega}+iI''_{\omega}$, with $I'_{\omega}$ and $I''_{\omega}$ expressing the in-phase and out-of-phase currents per electric field, respectively. 
When $\omega\rightarrow0$, we have $I''_{\|}=I''_{\perp}=0$.
The quantity $I_\omega$ is directly related to the conductance $G$ and conductivity $\sigma_0$ of the system. Assuming a uniform in-plane electric field, $I_\omega = G.d = \int_S\sigma_0dA$, where $d$ is the length of the sample along the direction of the applied electric field, and $S$ is the cross-sectional area.

\begin{figure}
    \centering
        \includegraphics[width=1.0\columnwidth]{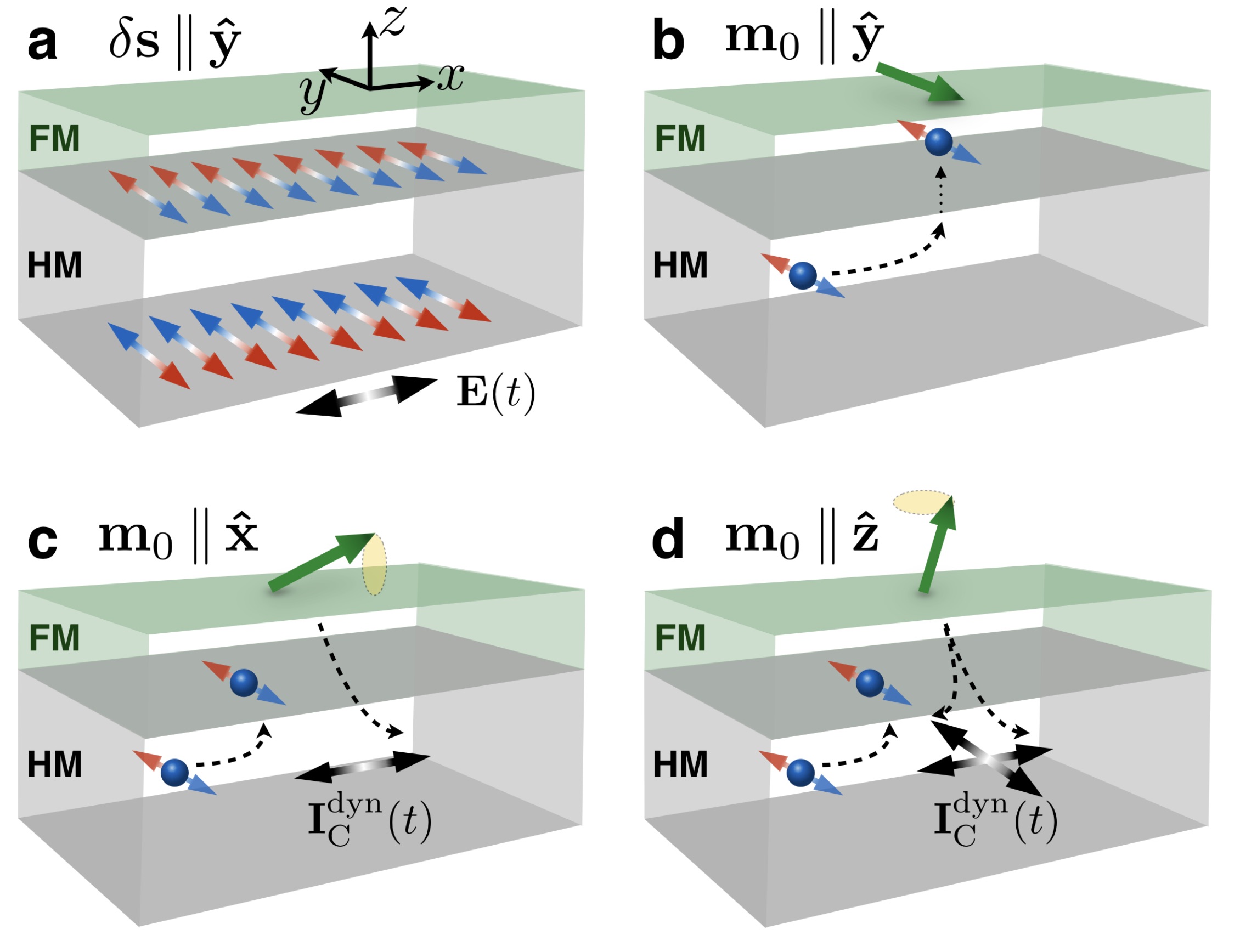}
    \caption{\textbf{Spin-orbit-related mechanisms in a ferromagnetic/heavy metal bilayer.} (a) 
    An ac electric field is applied to the whole sample, driving a charge current through the system. The current flowing through the heavy metal (HM) layer in turn generates an oscillatory spin accumulation $\delta\mathbf{s}\propto\mathbf{\hat{z}}\times\mathbf{E}(t)$ on its surfaces through spin Hall and inverse spin galvanic effects. (b-d) At the interface between the heavy metal and the ferromagnetic (FM) layer, $\delta\mathbf{s}$ interacts with its magnetization $\mathbf{m}_0$. (b) If $\mathbf{m}_0\,\|\,\delta\mathbf{s}$, the torque is zero and no magnetization dynamics is driven by the electric field. (c) In the case $\mathbf{m}_0$ is not aligned with $\mathbf{s}$, the magnetization can be set into precession, pumping charge and spins back into the heavy metal, thereby contributing to the current flowing across the system. (d) Depending on the direction of the magnetization, there can be contributions also to the transverse current.}
    \label{fig:diagram}
\end{figure}

A schematic setup and our choice of axes are illustrated in Fig.~\ref{fig:diagram}.
The total electric current flowing across the system can be split into two parts: $\mathbf{I}_\text{C}(t) = \mathbf{I}_\text{C}^0(t) + \mathbf{I}^\text{dyn}_\text{C}(t)$, where $\mathbf{I}^0_\text{C}(t)$ is the current driven directly by the electric field, and $\mathbf{I}^\text{dyn}_\text{C}(t)$ is the dynamical part produced by the magnetization precession. 
In the presence of spin-orbit coupling, $\mathbf{I}^0_\text{C}(t)$  depends upon the magnetization direction, giving rise to the anisotropic magnetoresistance (AMR) and also to transverse components due to the planar Hall effect (PHE) and anomalous Hall effect (AHE) \cite{McGuire:1975jk,Nagaosa:2010js,Seemann:2011fu,Bouaziz:2016fb}. 
Here, the spin Hall magnetoresistance (SMR) \cite{Nakayama:2013gs,Chen:2013kf} is also included in $\mathbf{I}^0_\text{C}(t)$. The frequency dependence of $\mathbf{I}^0_\text{C}(t)$ can be well described by the Drude model in metallic materials, and exhibits very little variation in the range of frequencies we are interested in. 

The dynamical contribution, $\mathbf{I}^\text{dyn}_\text{C}(t)$, originates from a combination of several mechanisms as follows: The spin Hall and inverse spin galvanic effects induce an oscillatory spin accumulation at the heavy metal surfaces, as illustrated on Fig.~\ref{fig:diagram}a. Then, spin-orbit torques may set the magnetization into precessional motion, depicted in Figs.~\ref{fig:diagram}(b-d). Note that when $\mathbf{m}_0\|\hat{\mathbf{y}}$, no precession is induced. Finally, spin pumping followed by inverse spin Hall effect and charge pumping generated by spin galvanic effects produce an additional charge current, as shown in Figs.~\ref{fig:diagram}(c,d). 


To reveal the spin-orbit origin of these effects, they are usually given in terms of the spin Hall angle $\theta_\text{SH}$. This quantity is defined as $\theta_{\text{SH}} \propto I_\perp^\text{S}/I_\|^\text{C}$, where $I_{\perp}^\text{S}$ is the transverse spin current with out-of-plane spin polarization and $I_{\|}^\text{C}$ is the longitudinal charge current.
It is a real number for dc currents, and measures how good the material is as a charge-to-spin converter. Phenomenologically, the symmetry of the dynamical contribution is similar to the SMR, and can be written as

\begin{equation}\label{eq:angular}
\begin{split}
\mathbf{I}^\text{dyn}_\text{C}(t) 
=&\left\{\Delta I_{\text{1}}(\omega)\left(m^{2}_0-m^{y2}_0\right)\mathbf{\hat{x}}+\Big[\Delta I_{\text{2}}(\omega)m^z_0+\Delta I_{\text{1}}(\omega)m^x_0m^y_0\Big]\mathbf{\hat{y}}\right\} E_0\cos(\omega t)
\end{split}
\end{equation}
This formula was obtained by considering the charge current converted from the pure spin current ($\propto\theta_{\text{SH}}$) that was pumped away from the magnetic layer due to the precessing magnetization \cite{{Tserkovnyak:2002ju}}. The motion of the magnetization is described by introducing the even and odd spin-orbit torques \cite{{Freimuth:2014kq}} (also $\propto\theta_{\text{SH}}$) into the Landau-Lifshitz-Gilbert equation. Therefore, $\Delta I_{\text{1,2}}(\omega)$ is proportional to $\theta_{\text{SH}}^2$, representing the two charge$\leftrightarrow$spin conversions involved in the effect (see Ref.~\citenum{{Tserkovnyak:2014jd}} and Supplementary Note S1). In this approach, $\theta_{\text{SH}}$ is assumed to be real.

The longitudinal and transverse contributions obtained in Eq.~2 are similar to the ones obtained for the SMR \cite{Chen:2013kf}. While the current flowing in the direction of the applied electric field ($\mathbf{\hat{x}}$) depends on the $y$ component of the magnetization (i.e., the direction of the spin accumulation induced by the spin-orbit coupling), the transverse current (flowing along the $\mathbf{\hat{y}}$ direction) has symmetries similar to the AHE ($\propto m_z$) and to the PHE ($\propto m_xm_y$). It is important to notice that, since this contribution involves the magnetization precession, $\Delta I_{1,2}$ are frequency-dependent. A detailed explanation of this contribution can be found in Supplementary Note S1.

To investigate the features of the dynamical effects described above, we use an ab-initio-based approach in which we map the electronic structure obtained in a density functional theory calculation into a multi-orbital tight-binding hamiltonian. 
The magnetic excitations and currents are obtained using linear response theory (see Methods and Supplementary Note S2). 

\subsection*{Dynamical currents and spin Hall angles}

We chose two bilayer metallic systems, Co/Pt(001) and Fe/W(110), manifesting different aspects of the phenomena due to the character of their charge to spin conversion.
There are two main differences between these systems. First, Co/Pt(001) presents an uniaxial perpendicular anisotropy, while Fe/W(110) has a biaxial in-plane anisotropy with the long axis being the lowest-energy direction. 
The second difference is the band filling of the substrate, which, according to Hund's rule, leads to parallel (anti-parallel) orbital and spin moments in Pt (W) due to the spin-orbit coupling \cite{Hoffmann:el}.
In our approach, the position of the resonance frequency is intrinsically determined by the electronic structure, without any adjustable parameter.
The magnetocrystalline anisotropy energies can be estimated through the modified Kittel's formula \cite{{Smit:1955wl},{Farle:1998gz}} or the static transverse magnetic susceptibility \cite{{dosSantosDias:2015bh}}. The values obtained using the latter and further details about the ground state properties of these systems are listed in Supplementary Note S3. 
The large anisotropies of the systems we investigate are mainly due to the fact that we are considering a single atomic ferromagnetic layer (See, for example, Refs.~\citenum{{Lehnert:2010fu},{Sipr:2010di}}). The strength and sign of the anisotropy energy can be tailored by changing the thickness and/or the composition of the two constituents forming the bilayer. For example, increasing the thickness of the FM layer lowers the effective anisotropy, which moves the resonance frequency to the GHz range.

Pt and W are known to have strong spin-orbit coupling and relatively high spin Hall angles. We have calculated $\theta_\text{SH}$ for slabs containing 10 layers of pure Pt(001) and pure W(110) and obtained the values 0.07 and -0.03, respectively, in accordance with experimental observations \cite{{RojasSanchez:2014ih},{Hoffmann:el}}. For a monolayer of Co deposited on Pt(001), we have found values of $\theta_\text{SH}\sim0.06$ for Pt thicknesses ranging from 4 to 9 atomic planes, in fair agreement with recent experimental estimates\cite{Garello:2014bi}.

\begin{figure*}
    \centering
        \includegraphics[width=1.0\textwidth]{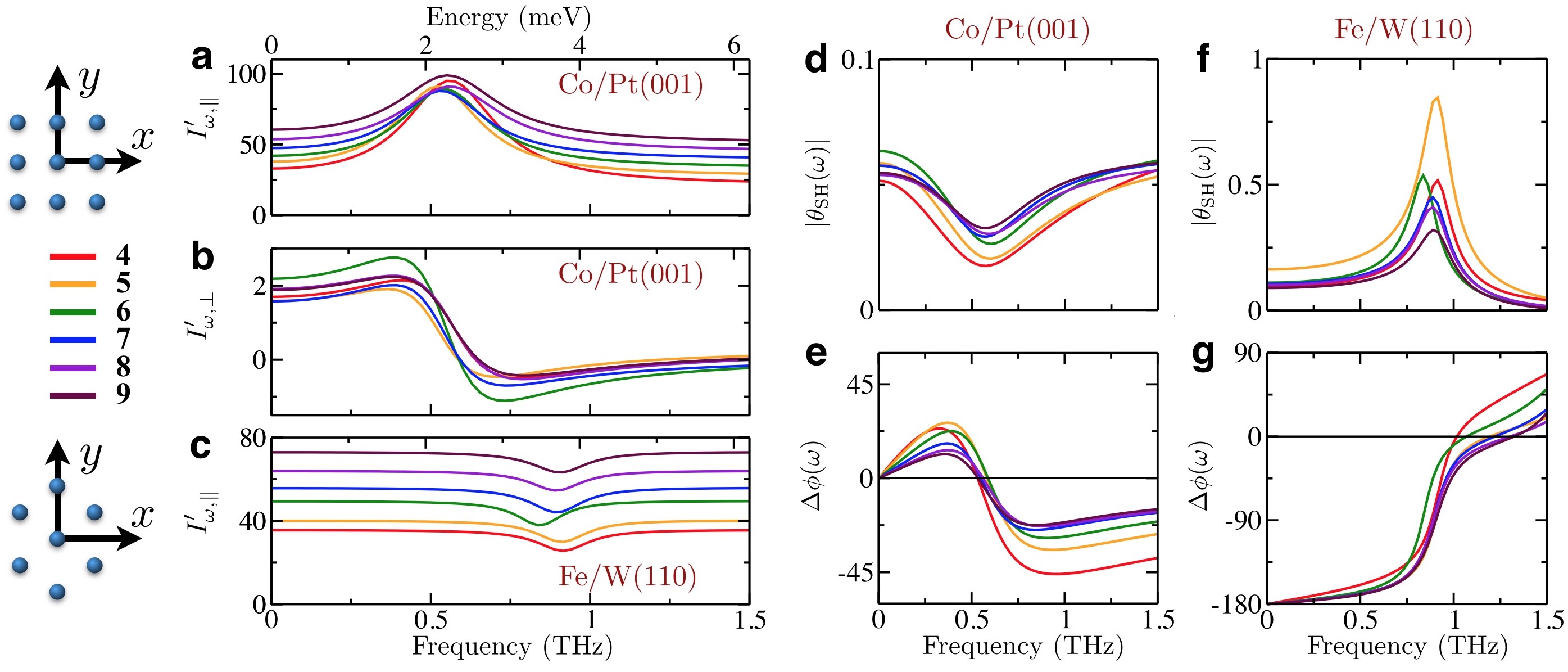}
    \caption{\textbf{Dependence of the longitudinal and transverse charge currents and of the dynamical spin Hall angle on the frequency of the applied electric field.} All calculations were performed for zero external magnetic field. (a) Longitudinal and (b) transverse in-phase currents calculated for monolayers of Co/Pt(001) as functions of the frequency, for different thicknesses of Pt. The magnetization is normal to the surface due to the easy axis anisotropy of the system. (c) Longitudinal in-phase currents calculated for monolayers of Fe/W(110) as functions of the frequency, for different W thicknesses. The magnetization lies in-plane in this case, again due to the magnetic anisotropy. As the electric field is parallel to the magnetization, $I'_{\omega,\perp}=0$ (not shown). (d-g) Amplitude and phase of the complex spin Hall angle, $\theta_\text{SH} = |\theta_\text{SH}|e^{i\Delta\phi}$, for Co/Pt(001) and Fe/W(110). The dc spin Hall angle is positive for Co/Pt(001) and negative for Fe/W(110), as seen from the phases. On the left side we indicate the substrate thickness, measured in atomic planes, and the position of the atoms on every layer with respect to our choice of axes for Co/Pt(001) (top) and Fe/W(110) (bottom).}
    \label{fig:thickness}
\end{figure*}

We start by exploring the behavior of the longitudinal and transverse currents, calculated as functions of the applied electric field frequency, for different thicknesses of the heavy metal substrate. The time-dependent electric field is applied along the [110] and [1$\bar{1}$0] directions for the Co/Pt(001) and Fe/W(110) systems, respectively, which defines the corresponding $x$ axis in each case. In Fig.~\ref{fig:thickness}, we plot the total first order in-phase currents along the directions parallel and perpendicular to the applied field, $I'_{\omega,\|}$ and $I'_{\omega,\perp}$, respectively. 
As expected from symmetry considerations, for Fe/W(110) with the electric field parallel to the magnetization direction, $I'_{\omega,\perp}=0$ (as illustrated in Fig.~\ref{fig:diagram}c).
A drastic variation, however, is observed for all the other currents as the frequency is varied. 
The structures manifested in the currents are related to the ferromagnetic resonances of the magnetic layers, located at frequencies determined by the magnetic anisotropy (calculations for Fig.~\ref{fig:thickness} were performed for zero external magnetic field). 
They induce substantial changes owing to the pumping mechanisms described earlier, which clearly affect both the longitudinal and transverse currents. 
We observe that the dynamical contributions for the longitudinal currents displayed in Figs.~\ref{fig:thickness}(a,c) exhibit symmetric line shapes in the vicinity of the resonance frequencies, whilst the line shapes of the transverse currents presented in Fig.~\ref{fig:thickness}b are clearly asymmetric. This can be explained by the properties of the charge-spin response function (see Supplementary Note S4). The thickness-dependent variations on the features shown in Fig.~\ref{fig:thickness} reflect quantum well effects present in our atomically thin films. These could be explored experimentally, e.g., by having ultrathin films of Pt or W deposited on an insulator.

Surprisingly, from the results of Figs.~\ref{fig:thickness}(a,c), we note that the longitudinal currents calculated for Co/Pt(001) exhibit a \textit{maximum}, whereas the corresponding ones for Fe/W(110) display a \textit{minimum}. 
This means that the \textit{dynamical} contribution to $I'_{\omega,\|}$ has opposite signs for Co/Pt(001) and Fe/W(110). 
Such a sign change has been experimentally observed in Ref.~\citenum{{Weiler:2014bp}} for permalloy deposited on Pt and Ta, and seems to contradict the phenomenological expectations from Eq.~\ref{eq:angular}. 
It turns out that the issue is not the postulated quadratic dependence on the spin Hall angle, but neglecting its dynamical nature: we must take into account the dephasing between the charge and spin currents, i.e., the spin Hall angle must be defined as a complex number. Figures~\ref{fig:thickness}(d,e) show the amplitude and phase of the spin Hall angle, respectively, obtained for Co/Pt(001), and Figs.~\ref{fig:thickness}(f,g) represents the same quantities for Fe/W(110). 
They reveal that the magnitude of the spin currents flowing perpendicularly to the electric field can be controlled by changing either its frequency or the strength of an applied external magnetic field (i.e., by tuning the ferromagnetic resonance frequency). 
On the one hand, for Co/Pt(001) the spin Hall angle amplitude decays to rather low values, close to turning off the flow of spin currents (see Fig.~\ref{fig:thickness}d). 
On the other hand, for Fe/W(110) it can reach giant values of $|\theta_\text{SH}|\sim0.5$ (see Fig.~\ref{fig:thickness}f) --- an enhancement of 500\% with respect to the dc value.
Complementary results displaying in-phase and out-of-phase spin Hall angles can be found in the Supplementary Figure S1.
We remark that the previous considerations on the importance of dephasing also applies to the other processes involved in the generation of the ac currents (such as the spin accumulation and the spin-orbit torques). 

An advantage of our approach is that all dephasing effects are naturally taken into account, and based on the underlying electronic structure.
Since these factors are also layer-dependent and include the oscillatory induced magnetic moments, they may give different contributions to the total currents.
For instance, our results show that increasing the thickness of the Pt(001) substrate leads to an increase of the current flowing through the Co layer, while the current on the Fe layer is quite insensitive on the thickness of W(110). 
The importance of how the current is distributed between the ferromagnetic and heavy metal layers has been raised in Refs.~\citenum{{Lau:2016hz},{Ghosh:2017ib}}.
Due to the non-homogeneous distribution of the currents along the layers, a time-dependent Oersted field can also be generated inside the sample. Nevertheless, in ultrathin films like the ones we investigate, it was shown and observed that this contribution is negligible compared to the spin-orbit torques \cite{{Miron:2011gd},{Garello:2013fa}, {Ciccarelli:2014bl}}.

\subsection*{Angular dependence of magnetoresistances and Hall effects}

The excitation of the spin waves depends on the relative angle between the magnetization and the induced spin accumulation at the interface of the bilayer. Hence, the angular dependence of the currents with respect to the magnetization direction --- which defines the magnetoresistance ratios and the Hall currents --- can be drastically modified.
To investigate this effect, we focus on bilayers consisting of a ferromagnetic monolayer deposited on four layers of heavy metal substrate. As in experimental setups, we consider a magnetic field larger than the anisotropy, to orient the magnetization along any chosen direction. 
Since increasing the static effective magnetic field diminishes the ability of the ac electric field to drive the magnetization precession, the dynamical charge currents are even stronger for smaller and more realistic external fields (see Supplementary Figure S2).

\begin{figure*}
    \centering
        \includegraphics[width=1.0\textwidth]{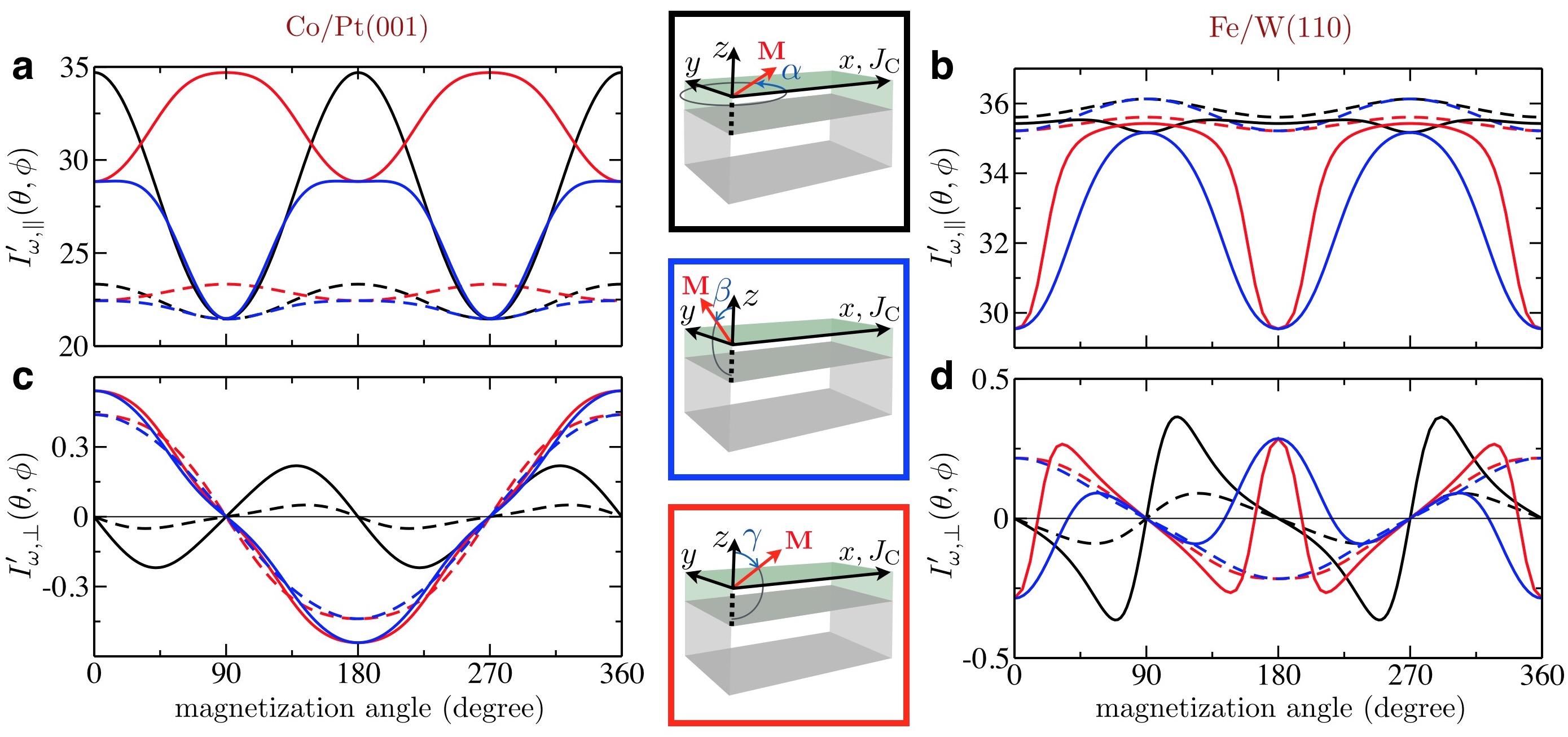}
    \caption{\textbf{Dependence of the dc and ac longitudinal and transverse charge currents on the direction of the magnetization.} 
    Longitudinal ($I'_{\omega,\|}$) and transverse ($I'_{\omega,\perp}$) in-phase currents, (a,c) for Co/Pt(001) and (b,d) for Fe/W(110). 
    The different rotation planes are illustrated on the middle panels: on the $xy$ plane (black, $\alpha$), $zy$ plane (blue, $\beta$) and $zx$ plane (red, $\gamma$). 
    The dashed curves correspond to an applied dc field: $I'_{\omega,\|}$ shows the AMR behavior for the $\gamma$ angular sweep (red) and the SMR behavior for the $\beta$ sweep (blue), while both AMR and SMR are present in the $\alpha$ sweep (black); $I'_{\omega,\perp}$ shows the PHE behavior for the $\alpha$ angular sweep (black) and the AHE behavior for the $\beta, \gamma$ sweeps (blue, red).
    Solid lines correspond to an applied ac field with the frequencies $\omega=\SI{1.3}{\tera\hertz}$ for Co/Pt(001) and $\omega=\SI{1.45}{\tera\hertz}$ for Fe/W(110), close to the ferromagnetic resonance frequency.
    The anisotropy in the transport properties is greatly enhanced and the angular dependence becomes non-trivial due to the variation in the amplitude and shape of the magnetization precession.
    }
    \label{fig:angular}
\end{figure*}

To demonstrate the dynamical effects on the magnetoresistances, we plot in Figs.~\ref{fig:angular}(a,b) the angular dependencies of the in-phase longitudinal currents for an applied dc electric field ($\omega=0$) and for a resonant ac field ($\omega=\omega_\text{res}$). 
In Table~\ref{tab:longitudinal}, we summarize the values of the variations $\Delta I^{\alpha,\beta,\gamma}=I'_{\omega,\|}(90^\circ)-I'_{\omega,\|}(0^\circ)$, where the angles $\alpha,\beta$ and $\gamma$ are illustrated on the central panel of Fig.~\ref{fig:angular}.
The large magnetoresistance enhancements (with respect to the static case) in the vicinities of the resonant frequencies are evident.
As expected from the results shown in the previous section, the average current increases for Co/Pt(001) and decreases for Fe/W(110).
Current variations that were almost vanishing in the dc limit (as the SMR and AMR changes depicted by the dashed blue curve in Fig.~\ref{fig:angular}a, and red curve in Fig.~\ref{fig:angular}b, respectively) become much larger than any static alterations.
It is worth noticing that the ac electric field can even change the sign of some of the magnetoresistance components.

The naive expectation is that the change in the currents is due to a modulation of the static magnetoresistance by the amplitude and shape of the magnetization precession \cite{Kim:2012bk}. 
This is true for frequencies away from the resonance, when the amplitude of precession and the pumped currents are small. 
Nevertheless, this is clearly not the case for the ac currents presented in Fig.~\ref{fig:angular}, otherwise the upper limit for the effect amplitude would be the dc magnetoresistance.

The curves displayed in Fig.~\ref{fig:angular} highlight the importance of higher order terms in the pumped currents, especially in the dynamical regime. 
Even though some of the obtained results for the $C_{4v}$ symmetry of Co/Pt(001) follow the behavior predicted by Eq.~\ref{eq:angular}, those for the $C_{2v}$ symmetry of Fe/W(110) definitely do not.
Large angular dependencies on the first order voltage were measured on strained (Ga,Mn)As, also $C_{2v}$\cite{Ciccarelli:2014bl}. However, they showed marked deviations with respect to predictions using phenomenological models.
These were attributed to the presence of higher order contributions on the pumped currents, which are intrinsically included in our results shown in Fig.~\ref{fig:angular}b.

As expected from the phenomenology, pumped currents also influence Hall effects.
In Figs.~\ref{fig:angular}(c,d), we show the in-phase currents that flow along the transverse direction, and we collect their maximum amplitude in Table~\ref{tab:transverse}.
In the dc-limit, we find that the PHE (dashed black curves) is smaller than the AHE (blue and red dashed curves) in both cases, as obtained in experimental measurements \cite{Garello:2013fa}. 
The dynamical contribution changes this picture and brings the maximum currents of all angular sweeps to the same order of magnitude.
The angular dependence presented in Fig.~\ref{fig:angular}d can be understood by analyzing the precessional motion of the magnetization (see Supplementary Note S4).

Other contributions not coming from the magnetization precession may also be relevant. 
As illustrated in Fig.~\ref{fig:diagram}b, spin waves are not excited when the magnetization points along the $y$ direction and, therefore, no dynamical contribution is expected.
However, a substantial change from dc to ac excitations can be seen on the black and blue curves of Fig.~\ref{fig:angular}b at $90^\circ$ or $270^\circ$.
We relate this effect to a possible precession of the spin current polarization around the magnetic field, an aspect not included in Fig.~\ref{fig:diagram}b and in Eq.~\ref{eq:angular}.

\begin{figure*}
    \centering
        \includegraphics[width=1.0\textwidth]{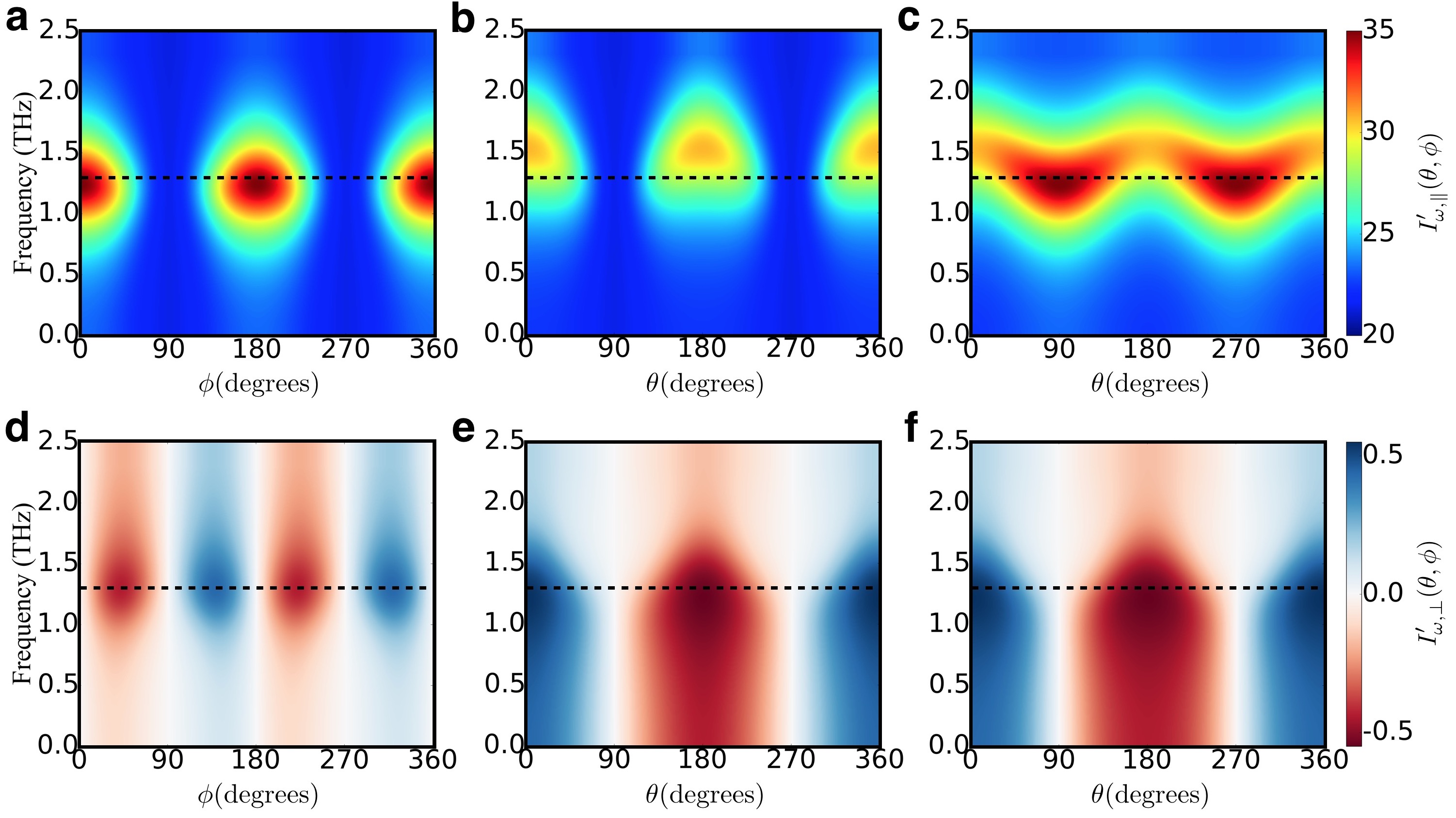}
    \caption{\textbf{Dependence of the ac charge currents on the direction of the magnetization and on the frequency of the applied electric field.} (a-c) Longitudinal currents and (d-f) transverse currents for Co/Pt(001), illustrated as a color map. 
    Different angular sweeps are shown: (a,d) xy plane; (b,e) zy plane and (c,f) zx plane. Results of panel d were multiplied by 2 for clarity.
    The dashed lines indicate the frequency used for the cuts depicted in Fig.~\ref{fig:angular}.
    The topography of the color map is determined by the crystallographic symmetry, the nature of the magnetic anisotropy and the applied magnetic field.
    Changing the frequency of the applied electric field gives control over the magnitude and shape of the dynamical magnetoresistances and Hall currents.
    }
    \label{fig:mr}
\end{figure*}

The angular variations discussed above depend on the chosen frequency of the applied field.
Figure~\ref{fig:mr} presents a complete map of the angular and frequency dependencies of all the dynamical effects described above, calculated for a monolayer of Co on 4 layers of Pt(001). 
The black dashed lines indicate the frequency in which the angular dependencies of the dynamical currents plotted in Figs.~\ref{fig:angular}(a,c) were obtained.
They provide a clear picture of how the precessional motion of the magnetization influences longitudinal and charge currents.
Different magnetoresistances and Hall currents can be obtained depending on the anisotropy of the system and the applied magnetic fields. 
Therefore, by tuning the frequency of the oscillatory field to a suitable value, one can obtain large current variations as a result of slight changes in the magnetization direction.

\section*{Discussion}

We have investigated the dynamical effects on the linear response charge currents due to spin-orbit-related phenomena. 
They involve the simultaneous action of both spin-Hall and inverse spin galvanic effects together with their reciprocal counterparts, as in the SMR, with an additional contribution arising from the excitation of the ferromagnetic resonance by the ac spin-orbit torques. 
We have presented results for Co/Pt(001) and Fe/W(110) that unveil the complex nature of spin-charge excitations in these systems. 
They show that relevant physical mechanisms are being neglected by current phenomenological approaches, which consequently fail to properly describe the amplitude and angular dependence of the ac currents induced in such setups.
Our material-dependent first-principles approach avoid these pitfalls by taking into account the lattice structure and all the dephasing factors between different physical quantities, which are intrinsically determined by the electronic structure.
Although our main focus is metallic multilayers, the underlying processes discussed here also apply to other systems such as magnetic insulators in contact with heavy metals (e.g. YIG/Pt), and magnetic semiconductors with relatively large SOC, where the spin pumping contribution to the magnetoresistance can be also relevant \cite{Ciccarelli:2014bl}.

We propose that the dynamical contributions, that are present not only in the charge but also in the spin currents, can be used to manipulate the spin Hall angle of the system. 
This means that the flow of spin currents can be controlled by changing either the frequency of the electric field or the intensity of the magnetic field. 
Recently, giant spin Hall angles were measured for a few systems in the dc limit \cite{{Liu:2012dga},{Pai:2012ef},{Zhang:2016bf}}. Our work indicates a huge potential for magnifying even further these angles in the ac regime.


The dynamical processes we describe manifest prominently in the ac currents flowing with the same frequency as the external perturbation (first harmonic signal). 
Nevertheless, the frequency and field dependence of the current response may also affect dc and second harmonic signals, especially for frequencies close to the ferromagnetic resonance of the system (see Supplementary Note S5).
This may be related to why the second harmonic signal, measured in the quasi-static limit, provides more reliable results than the dc measurements (see Supplemental Material of Ref.~\citenum{Garello:2013fa}). 

Since the main contribution to the charge current is generated by the magnetization precession, it can be deactivated by rotating the magnetization direction, substantially modifying dynamical magnetoresistances.
Hall currents can also be sensitively controlled by tuning the frequency of the applied electric field in the vicinity of the ferromagnetic resonance.
These mechanisms can be used to engineer drastic variations of the ac currents upon rotation of the magnetization by only a few degrees.
Conversely, ac currents can be employed for manipulating magnetic states due to the enhancement of the spin-orbit torque, in particular at the ferromagnetic resonance frequency. 
This change in current signals provides an efficient way of manipulating and detecting magnetic states and may be an alternative to the microwave-assisted magnetic recording. A hypothetical protocol exploiting these effects could be: first apply a short weak current pulse that sets the magnetization into precessional motion, immediately followed by a larger pulse that would induce the switching.
Current pulses may be used to incorporate high frequency components into the actuating current

The control of the current signal may help to unravel mechanisms to improve and measure spin-orbit torques, providing an efficient way to manipulate magnetic states, e.g., through microwave assisted recording technologies \cite{Okamoto:2015gq}.
The dynamical physical concepts described above can be adjusted by choosing lattice symmetries, material elements and layer thicknesses of multilayer structures. 
This engineering procedure already proved effective in building efficient THz devices by taking advantage of the different spin-orbit properties of Pt and W \cite{SeifertT:2016kc}.
Our results demonstrate that the exploration of dynamical contribution gives an extra degree of freedom to manipulate future spintronic devices and may support ultrafast reading and writing mechanisms. 

\section*{Methods}

The electronic structures of the adsorbate/substrate combinations are described by a multi-orbital tight-binding hamiltonian with nine orbitals (the five $d$ and the $sp$ complex) and the two spin components included per atom. The hopping matrices are obtained from density functional calculations based on the real-space linear-muffin tin orbitals method, within the atomic sphere approximation (RS-LMTO-ASA) \cite{{Andersen:1984ir},{Peduto:1991hn},{FrotaPessoa:1992kt}}. Mean field theory is used to describe the ground state of the system, considering an effective intra-atomic Coulomb interaction $U=\SI{1}{\electronvolt}$ within the d shells of both the ferromagnetic and heavy metal layers. The ground-state magnetization of the ferromagnetic layer is calculated self-consistently, assuming a fixed value of the Fermi energy and adjusting the atomic $d$-orbital energy levels to reproduce the electronic occupancy obtained by the RS-LMTO-ASA calculations for each layer \cite{bmcm}. At the end of this process the atomic planes of the non-magnetic substrate close to ferromagnetic layer become also slightly spin polarized due to the proximity effect. 
Spin-orbit interaction is incorporated within the local (on-site) approximation $\hat{H}_\text{SO} = \sum_i \lambda_i\mathbf{L}_i\cdot\mathbf{S}_i$, explicitly involving the $d$-band complex only. All coupling constants $\lambda_i$ are obtained from RS-LMTO-ASA first principles calculations. Spin rotational symmetry is broken in the presence of SOC, and in the cases we consider, it leads to a perpendicular magnetic anisotropy for the Co/Pt(001) system and the [1$\bar{1}$0] direction being the easy axis for Fe/W(110).

Our description of the excitations is based on linear response theory using the Kubo formalism \cite{Kubo:1957cl}, and follows through application of the RPA. Due to spin-orbit coupling, the responses of the transverse components of the magnetization are coupled to the longitudinal ones, and we end up with a $4\times4$ susceptibility matrix in spin space whose elements need to be calculated simultaneously \cite{{Costa:2006hb},{Costa2010}}. 

A more detailed description of the theory is given in the Supplementary Note S2.

%
%
%

\bibliography{ac-SMR.bib}

\begin{thebibliography}{54}%
\makeatletter
\providecommand \@ifxundefined [1]{%
 \@ifx{#1\undefined}
}%
\providecommand \@ifnum [1]{%
 \ifnum #1\expandafter \@firstoftwo
 \else \expandafter \@secondoftwo
 \fi
}%
\providecommand \@ifx [1]{%
 \ifx #1\expandafter \@firstoftwo
 \else \expandafter \@secondoftwo
 \fi
}%
\providecommand \natexlab [1]{#1}%
\providecommand \enquote  [1]{``#1''}%
\providecommand \bibnamefont  [1]{#1}%
\providecommand \bibfnamefont [1]{#1}%
\providecommand \citenamefont [1]{#1}%
\providecommand \href@noop [0]{\@secondoftwo}%
\providecommand \href [0]{\begingroup \@sanitize@url \@href}%
\providecommand \@href[1]{\@@startlink{#1}\@@href}%
\providecommand \@@href[1]{\endgroup#1\@@endlink}%
\providecommand \@sanitize@url [0]{\catcode `\\12\catcode `\$12\catcode
  `\&12\catcode `\#12\catcode `\^12\catcode `\_12\catcode `\%12\relax}%
\providecommand \@@startlink[1]{}%
\providecommand \@@endlink[0]{}%
\providecommand \url  [0]{\begingroup\@sanitize@url \@url }%
\providecommand \@url [1]{\endgroup\@href {#1}{\urlprefix }}%
\providecommand \urlprefix  [0]{URL }%
\providecommand \Eprint [0]{\href }%
\providecommand \doibase [0]{http://dx.doi.org/}%
\providecommand \selectlanguage [0]{\@gobble}%
\providecommand \bibinfo  [0]{\@secondoftwo}%
\providecommand \bibfield  [0]{\@secondoftwo}%
\providecommand \translation [1]{[#1]}%
\providecommand \BibitemOpen [0]{}%
\providecommand \bibitemStop [0]{}%
\providecommand \bibitemNoStop [0]{.\EOS\space}%
\providecommand \EOS [0]{\spacefactor3000\relax}%
\providecommand \BibitemShut  [1]{\csname bibitem#1\endcsname}%
\let\auto@bib@innerbib\@empty
\bibitem [{\citenamefont {Thomson}(1856)}]{Thomson:1856jw}%
  \BibitemOpen
  \bibfield  {author} {\bibinfo {author} {\bibfnamefont {W}~\bibnamefont
  {Thomson}},\ }\bibfield  {title} {\enquote {\bibinfo {title} {{On the
  Electro-Dynamic Qualities of Metals:--Effects of Magnetization on the
  Electric Conductivity of Nickel and of Iron}},}\ }\href@noop {} {\bibfield
  {journal} {\bibinfo  {journal} {Proc. R. Soc. Lond.}\ }\textbf {\bibinfo
  {volume} {8}},\ \bibinfo {pages} {546--550} (\bibinfo {year}
  {1856})}\BibitemShut {NoStop}%
\bibitem [{\citenamefont {Jan}(1957)}]{Jan:1957gk}%
  \BibitemOpen
  \bibfield  {author} {\bibinfo {author} {\bibfnamefont {J~P}\ \bibnamefont
  {Jan}},\ }\bibfield  {title} {\enquote {\bibinfo {title} {{Galvamomagnetic
  and Thermomagnetic Effects in Metals}},}\ }\href@noop {} {\bibfield
  {journal} {\bibinfo  {journal} {Solid State Physics}\ }\textbf {\bibinfo
  {volume} {5}},\ \bibinfo {pages} {1--96} (\bibinfo {year}
  {1957})}\BibitemShut {NoStop}%
\bibitem [{\citenamefont {Nakayama}\ \emph {et~al.}(2013)\citenamefont
  {Nakayama}, \citenamefont {Althammer}, \citenamefont {Chen}, \citenamefont
  {Uchida}, \citenamefont {Kajiwara}, \citenamefont {Kikuchi}, \citenamefont
  {Ohtani}, \citenamefont {Gepr{\"a}gs}, \citenamefont {Opel}, \citenamefont
  {Takahashi}, \citenamefont {Gross}, \citenamefont {Bauer}, \citenamefont
  {Goennenwein},\ and\ \citenamefont {Saitoh}}]{Nakayama:2013gs}%
  \BibitemOpen
  \bibfield  {author} {\bibinfo {author} {\bibfnamefont {H}~\bibnamefont
  {Nakayama}}, \bibinfo {author} {\bibfnamefont {M}~\bibnamefont {Althammer}},
  \bibinfo {author} {\bibfnamefont {Y~T}\ \bibnamefont {Chen}}, \bibinfo
  {author} {\bibfnamefont {K}~\bibnamefont {Uchida}}, \bibinfo {author}
  {\bibfnamefont {Y.}~\bibnamefont {Kajiwara}}, \bibinfo {author}
  {\bibfnamefont {D}~\bibnamefont {Kikuchi}}, \bibinfo {author} {\bibfnamefont
  {T}~\bibnamefont {Ohtani}}, \bibinfo {author} {\bibfnamefont {S}~\bibnamefont
  {Gepr{\"a}gs}}, \bibinfo {author} {\bibfnamefont {M}~\bibnamefont {Opel}},
  \bibinfo {author} {\bibfnamefont {S.}~\bibnamefont {Takahashi}}, \bibinfo
  {author} {\bibfnamefont {R}~\bibnamefont {Gross}}, \bibinfo {author}
  {\bibfnamefont {G.~E.~W.}\ \bibnamefont {Bauer}}, \bibinfo {author}
  {\bibfnamefont {S~T~B}\ \bibnamefont {Goennenwein}}, \ and\ \bibinfo {author}
  {\bibfnamefont {E.}~\bibnamefont {Saitoh}},\ }\bibfield  {title} {\enquote
  {\bibinfo {title} {{Spin Hall Magnetoresistance Induced by a Nonequilibrium
  Proximity Effect}},}\ }\href@noop {} {\bibfield  {journal} {\bibinfo
  {journal} {Phys. Rev. Lett.}\ }\textbf {\bibinfo {volume} {110}},\ \bibinfo
  {pages} {206601} (\bibinfo {year} {2013})}\BibitemShut {NoStop}%
\bibitem [{\citenamefont {Chen}\ \emph {et~al.}(2013)\citenamefont {Chen},
  \citenamefont {Takahashi}, \citenamefont {Nakayama}, \citenamefont
  {Althammer}, \citenamefont {Goennenwein}, \citenamefont {Saitoh},\ and\
  \citenamefont {Bauer}}]{Chen:2013kf}%
  \BibitemOpen
  \bibfield  {author} {\bibinfo {author} {\bibfnamefont {Yan-Ting}\
  \bibnamefont {Chen}}, \bibinfo {author} {\bibfnamefont {S.}~\bibnamefont
  {Takahashi}}, \bibinfo {author} {\bibfnamefont {Hiroyasu}\ \bibnamefont
  {Nakayama}}, \bibinfo {author} {\bibfnamefont {Matthias}\ \bibnamefont
  {Althammer}}, \bibinfo {author} {\bibfnamefont {Sebastian T~B}\ \bibnamefont
  {Goennenwein}}, \bibinfo {author} {\bibfnamefont {E.}~\bibnamefont {Saitoh}},
  \ and\ \bibinfo {author} {\bibfnamefont {G.~E.~W.}\ \bibnamefont {Bauer}},\
  }\bibfield  {title} {\enquote {\bibinfo {title} {{Theory of spin Hall
  magnetoresistance}},}\ }\href@noop {} {\bibfield  {journal} {\bibinfo
  {journal} {Phys. Rev. B}\ }\textbf {\bibinfo {volume} {87}},\ \bibinfo
  {pages} {144411} (\bibinfo {year} {2013})}\BibitemShut {NoStop}%
\bibitem [{\citenamefont {Avci}\ \emph {et~al.}(2015)\citenamefont {Avci},
  \citenamefont {Garello}, \citenamefont {Ghosh}, \citenamefont {Gabureac},
  \citenamefont {Alvarado},\ and\ \citenamefont {Gambardella}}]{Avci:2015jp}%
  \BibitemOpen
  \bibfield  {author} {\bibinfo {author} {\bibfnamefont {C.~O.}\ \bibnamefont
  {Avci}}, \bibinfo {author} {\bibfnamefont {K.}~\bibnamefont {Garello}},
  \bibinfo {author} {\bibfnamefont {A.}~\bibnamefont {Ghosh}}, \bibinfo
  {author} {\bibfnamefont {M.}~\bibnamefont {Gabureac}}, \bibinfo {author}
  {\bibfnamefont {S.~F.}\ \bibnamefont {Alvarado}}, \ and\ \bibinfo {author}
  {\bibfnamefont {P.}~\bibnamefont {Gambardella}},\ }\bibfield  {title}
  {\enquote {\bibinfo {title} {{Unidirectional spin Hall magnetoresistance in
  ferromagnet/normal metal bilayers}},}\ }\href@noop {} {\bibfield  {journal}
  {\bibinfo  {journal} {Nature Phys.}\ }\textbf {\bibinfo {volume} {11}},\
  \bibinfo {pages} {570--575} (\bibinfo {year} {2015})}\BibitemShut {NoStop}%
\bibitem [{\citenamefont {Olejn{\'\i}k}\ \emph {et~al.}(2015)\citenamefont
  {Olejn{\'\i}k}, \citenamefont {Nov{\'a}k}, \citenamefont {Wunderlich},\ and\
  \citenamefont {Jungwirth}}]{Olejnik:2015bc}%
  \BibitemOpen
  \bibfield  {author} {\bibinfo {author} {\bibfnamefont {K.}~\bibnamefont
  {Olejn{\'\i}k}}, \bibinfo {author} {\bibfnamefont {V.}~\bibnamefont
  {Nov{\'a}k}}, \bibinfo {author} {\bibfnamefont {J.}~\bibnamefont
  {Wunderlich}}, \ and\ \bibinfo {author} {\bibfnamefont {T.}~\bibnamefont
  {Jungwirth}},\ }\bibfield  {title} {\enquote {\bibinfo {title} {{Electrical
  detection of magnetization reversal without auxiliary magnets}},}\
  }\href@noop {} {\bibfield  {journal} {\bibinfo  {journal} {Phys. Rev. B}\
  }\textbf {\bibinfo {volume} {91}},\ \bibinfo {pages} {180402} (\bibinfo
  {year} {2015})}\BibitemShut {NoStop}%
\bibitem [{\citenamefont {V{\'e}lez}\ \emph {et~al.}(2016)\citenamefont
  {V{\'e}lez}, \citenamefont {Golovach}, \citenamefont {Bedoya-Pinto},
  \citenamefont {Isasa}, \citenamefont {Sagasta}, \citenamefont {Abadia},
  \citenamefont {Rogero}, \citenamefont {Hueso}, \citenamefont {Bergeret},\
  and\ \citenamefont {Casanova}}]{Velez:2016bm}%
  \BibitemOpen
  \bibfield  {author} {\bibinfo {author} {\bibfnamefont {Sa{\"u}l}\
  \bibnamefont {V{\'e}lez}}, \bibinfo {author} {\bibfnamefont {Vitaly~N}\
  \bibnamefont {Golovach}}, \bibinfo {author} {\bibfnamefont {Amilcar}\
  \bibnamefont {Bedoya-Pinto}}, \bibinfo {author} {\bibfnamefont {Miren}\
  \bibnamefont {Isasa}}, \bibinfo {author} {\bibfnamefont {Edurne}\
  \bibnamefont {Sagasta}}, \bibinfo {author} {\bibfnamefont {Mikel}\
  \bibnamefont {Abadia}}, \bibinfo {author} {\bibfnamefont {Celia}\
  \bibnamefont {Rogero}}, \bibinfo {author} {\bibfnamefont {Luis~E}\
  \bibnamefont {Hueso}}, \bibinfo {author} {\bibfnamefont {F~Sebastian}\
  \bibnamefont {Bergeret}}, \ and\ \bibinfo {author} {\bibfnamefont
  {F{\`e}lix}\ \bibnamefont {Casanova}},\ }\bibfield  {title} {\enquote
  {\bibinfo {title} {{Hanle Magnetoresistance in Thin Metal Films with Strong
  Spin-Orbit Coupling}},}\ }\href@noop {} {\bibfield  {journal} {\bibinfo
  {journal} {Phys. Rev. Lett.}\ }\textbf {\bibinfo {volume} {116}},\ \bibinfo
  {pages} {016603} (\bibinfo {year} {2016})}\BibitemShut {NoStop}%
\bibitem [{\citenamefont {Nakayama}\ \emph {et~al.}(2016)\citenamefont
  {Nakayama}, \citenamefont {Kanno}, \citenamefont {An}, \citenamefont
  {Tashiro}, \citenamefont {Haku}, \citenamefont {Nomura},\ and\ \citenamefont
  {Ando}}]{Nakayama:2016gp}%
  \BibitemOpen
  \bibfield  {author} {\bibinfo {author} {\bibfnamefont {Hiroyasu}\
  \bibnamefont {Nakayama}}, \bibinfo {author} {\bibfnamefont {Yusuke}\
  \bibnamefont {Kanno}}, \bibinfo {author} {\bibfnamefont {Hongyu}\
  \bibnamefont {An}}, \bibinfo {author} {\bibfnamefont {Takaharu}\ \bibnamefont
  {Tashiro}}, \bibinfo {author} {\bibfnamefont {Satoshi}\ \bibnamefont {Haku}},
  \bibinfo {author} {\bibfnamefont {Akiyo}\ \bibnamefont {Nomura}}, \ and\
  \bibinfo {author} {\bibfnamefont {Kazuya}\ \bibnamefont {Ando}},\ }\bibfield
  {title} {\enquote {\bibinfo {title} {{Rashba-Edelstein Magnetoresistance in
  Metallic Heterostructures}},}\ }\href@noop {} {\bibfield  {journal} {\bibinfo
   {journal} {Phys. Rev. Lett.}\ }\textbf {\bibinfo {volume} {117}},\ \bibinfo
  {pages} {116602} (\bibinfo {year} {2016})}\BibitemShut {NoStop}%
\bibitem [{\citenamefont {Fang}\ \emph {et~al.}(2011)\citenamefont {Fang},
  \citenamefont {Kurebayashi}, \citenamefont {Wunderlich}, \citenamefont
  {V{\'{y}}born{\'{y}}}, \citenamefont {Z{\^a}rbo}, \citenamefont {Campion},
  \citenamefont {Casiraghi}, \citenamefont {Gallagher}, \citenamefont
  {Jungwirth},\ and\ \citenamefont {Ferguson}}]{Fang:2011ch}%
  \BibitemOpen
  \bibfield  {author} {\bibinfo {author} {\bibfnamefont {D.}~\bibnamefont
  {Fang}}, \bibinfo {author} {\bibfnamefont {H.}~\bibnamefont {Kurebayashi}},
  \bibinfo {author} {\bibfnamefont {J.}~\bibnamefont {Wunderlich}}, \bibinfo
  {author} {\bibfnamefont {K.}~\bibnamefont {V{\'{y}}born{\'{y}}}}, \bibinfo
  {author} {\bibfnamefont {L.~P.}\ \bibnamefont {Z{\^a}rbo}}, \bibinfo {author}
  {\bibfnamefont {R.~P.}\ \bibnamefont {Campion}}, \bibinfo {author}
  {\bibfnamefont {A.}~\bibnamefont {Casiraghi}}, \bibinfo {author}
  {\bibfnamefont {B.~L.}\ \bibnamefont {Gallagher}}, \bibinfo {author}
  {\bibfnamefont {T.}~\bibnamefont {Jungwirth}}, \ and\ \bibinfo {author}
  {\bibfnamefont {A.~J.}\ \bibnamefont {Ferguson}},\ }\bibfield  {title}
  {\enquote {\bibinfo {title} {{Spin{\textendash}orbit-driven ferromagnetic
  resonance}},}\ }\href@noop {} {\bibfield  {journal} {\bibinfo  {journal}
  {Nature Nanotechnol.}\ }\textbf {\bibinfo {volume} {6}},\ \bibinfo {pages}
  {413--417} (\bibinfo {year} {2011})}\BibitemShut {NoStop}%
\bibitem [{\citenamefont {Garello}\ \emph {et~al.}(2013)\citenamefont
  {Garello}, \citenamefont {Miron}, \citenamefont {Avci}, \citenamefont
  {Freimuth}, \citenamefont {Mokrousov}, \citenamefont {Bl{\"u}gel},
  \citenamefont {Auffret}, \citenamefont {Boulle}, \citenamefont {Gaudin},\
  and\ \citenamefont {Gambardella}}]{Garello:2013fa}%
  \BibitemOpen
  \bibfield  {author} {\bibinfo {author} {\bibfnamefont {K.}~\bibnamefont
  {Garello}}, \bibinfo {author} {\bibfnamefont {I.~M.}\ \bibnamefont {Miron}},
  \bibinfo {author} {\bibfnamefont {C.~O.}\ \bibnamefont {Avci}}, \bibinfo
  {author} {\bibfnamefont {F.}~\bibnamefont {Freimuth}}, \bibinfo {author}
  {\bibfnamefont {Y.}~\bibnamefont {Mokrousov}}, \bibinfo {author}
  {\bibfnamefont {S.}~\bibnamefont {Bl{\"u}gel}}, \bibinfo {author}
  {\bibfnamefont {S.}~\bibnamefont {Auffret}}, \bibinfo {author} {\bibfnamefont
  {O.}~\bibnamefont {Boulle}}, \bibinfo {author} {\bibfnamefont
  {G.}~\bibnamefont {Gaudin}}, \ and\ \bibinfo {author} {\bibfnamefont
  {P.}~\bibnamefont {Gambardella}},\ }\bibfield  {title} {\enquote {\bibinfo
  {title} {{Symmetry and magnitude of spin-orbit torques in ferromagnetic
  heterostructures}},}\ }\href@noop {} {\bibfield  {journal} {\bibinfo
  {journal} {Nature Nanotechnol.}\ }\textbf {\bibinfo {volume} {8}},\ \bibinfo
  {pages} {587--593} (\bibinfo {year} {2013})}\BibitemShut {NoStop}%
\bibitem [{\citenamefont {Freimuth}\ \emph {et~al.}(2014)\citenamefont
  {Freimuth}, \citenamefont {Bl{\"u}gel},\ and\ \citenamefont
  {Mokrousov}}]{Freimuth:2014kq}%
  \BibitemOpen
  \bibfield  {author} {\bibinfo {author} {\bibfnamefont {F.}~\bibnamefont
  {Freimuth}}, \bibinfo {author} {\bibfnamefont {S.}~\bibnamefont
  {Bl{\"u}gel}}, \ and\ \bibinfo {author} {\bibfnamefont {Y.}~\bibnamefont
  {Mokrousov}},\ }\bibfield  {title} {\enquote {\bibinfo {title} {{Spin-orbit
  torques in Co/Pt(111) and Mn/W(001) magnetic bilayers from first
  principles}},}\ }\href@noop {} {\bibfield  {journal} {\bibinfo  {journal}
  {Phys. Rev. B}\ }\textbf {\bibinfo {volume} {90}},\ \bibinfo {pages} {174423}
  (\bibinfo {year} {2014})}\BibitemShut {NoStop}%
\bibitem [{\citenamefont {Freimuth}\ \emph {et~al.}(2015)\citenamefont
  {Freimuth}, \citenamefont {Bl{\"u}gel},\ and\ \citenamefont
  {Mokrousov}}]{Freimuth:2015fq}%
  \BibitemOpen
  \bibfield  {author} {\bibinfo {author} {\bibfnamefont {F.}~\bibnamefont
  {Freimuth}}, \bibinfo {author} {\bibfnamefont {S.}~\bibnamefont
  {Bl{\"u}gel}}, \ and\ \bibinfo {author} {\bibfnamefont {Y.}~\bibnamefont
  {Mokrousov}},\ }\bibfield  {title} {\enquote {\bibinfo {title} {{Direct and
  inverse spin-orbit torques}},}\ }\href@noop {} {\bibfield  {journal}
  {\bibinfo  {journal} {Phys. Rev. B}\ }\textbf {\bibinfo {volume} {92}},\
  \bibinfo {pages} {064415} (\bibinfo {year} {2015})}\BibitemShut {NoStop}%
\bibitem [{\citenamefont {Guimar{\~a}es}\ \emph {et~al.}(2015)\citenamefont
  {Guimar{\~a}es}, \citenamefont {Lounis}, \citenamefont {Costa},\ and\
  \citenamefont {Muniz}}]{Guimaraes:2015fl}%
  \BibitemOpen
  \bibfield  {author} {\bibinfo {author} {\bibfnamefont {F.~S.~M.}\
  \bibnamefont {Guimar{\~a}es}}, \bibinfo {author} {\bibfnamefont
  {S.}~\bibnamefont {Lounis}}, \bibinfo {author} {\bibfnamefont {A.~T.}\
  \bibnamefont {Costa}}, \ and\ \bibinfo {author} {\bibfnamefont {R.~B.}\
  \bibnamefont {Muniz}},\ }\bibfield  {title} {\enquote {\bibinfo {title}
  {{Dynamical current-induced ferromagnetic and antiferromagnetic
  resonances}},}\ }\href@noop {} {\bibfield  {journal} {\bibinfo  {journal}
  {Phys. Rev. B}\ }\textbf {\bibinfo {volume} {92}},\ \bibinfo {pages}
  {220410(R)} (\bibinfo {year} {2015})}\BibitemShut {NoStop}%
\bibitem [{\citenamefont {Miron}\ \emph {et~al.}(2011)\citenamefont {Miron},
  \citenamefont {Garello}, \citenamefont {Gaudin}, \citenamefont {Zermatten},
  \citenamefont {Costache}, \citenamefont {Auffret}, \citenamefont {Bandiera},
  \citenamefont {Rodmacq}, \citenamefont {Schuhl},\ and\ \citenamefont
  {Gambardella}}]{Miron:2011gd}%
  \BibitemOpen
  \bibfield  {author} {\bibinfo {author} {\bibfnamefont {I.~M.}\ \bibnamefont
  {Miron}}, \bibinfo {author} {\bibfnamefont {K.}~\bibnamefont {Garello}},
  \bibinfo {author} {\bibfnamefont {G.}~\bibnamefont {Gaudin}}, \bibinfo
  {author} {\bibfnamefont {Pierre-Jean}\ \bibnamefont {Zermatten}}, \bibinfo
  {author} {\bibfnamefont {Marius~V}\ \bibnamefont {Costache}}, \bibinfo
  {author} {\bibfnamefont {S.}~\bibnamefont {Auffret}}, \bibinfo {author}
  {\bibfnamefont {S{\'e}bastien}\ \bibnamefont {Bandiera}}, \bibinfo {author}
  {\bibfnamefont {Bernard}\ \bibnamefont {Rodmacq}}, \bibinfo {author}
  {\bibfnamefont {Alain}\ \bibnamefont {Schuhl}}, \ and\ \bibinfo {author}
  {\bibfnamefont {Pietro}\ \bibnamefont {Gambardella}},\ }\bibfield  {title}
  {\enquote {\bibinfo {title} {{Perpendicular switching of a single
  ferromagnetic layer induced by in-plane current injection}},}\ }\href@noop {}
  {\bibfield  {journal} {\bibinfo  {journal} {Nat. Commun.}\ }\textbf {\bibinfo
  {volume} {476}},\ \bibinfo {pages} {189--193} (\bibinfo {year}
  {2011})}\BibitemShut {NoStop}%
\bibitem [{\citenamefont {Pai}\ \emph {et~al.}(2012)\citenamefont {Pai},
  \citenamefont {Liu}, \citenamefont {Li}, \citenamefont {Tseng}, \citenamefont
  {Ralph},\ and\ \citenamefont {Buhrman}}]{Pai:2012ef}%
  \BibitemOpen
  \bibfield  {author} {\bibinfo {author} {\bibfnamefont {Chi-Feng}\
  \bibnamefont {Pai}}, \bibinfo {author} {\bibfnamefont {L.}~\bibnamefont
  {Liu}}, \bibinfo {author} {\bibfnamefont {Y}~\bibnamefont {Li}}, \bibinfo
  {author} {\bibfnamefont {H~W}\ \bibnamefont {Tseng}}, \bibinfo {author}
  {\bibfnamefont {D.~C.}\ \bibnamefont {Ralph}}, \ and\ \bibinfo {author}
  {\bibfnamefont {R.~A.}\ \bibnamefont {Buhrman}},\ }\bibfield  {title}
  {\enquote {\bibinfo {title} {{Spin transfer torque devices utilizing the
  giant spin Hall effect of tungsten}},}\ }\href@noop {} {\bibfield  {journal}
  {\bibinfo  {journal} {Appl. Phys. Lett.}\ }\textbf {\bibinfo {volume}
  {101}},\ \bibinfo {pages} {122404} (\bibinfo {year} {2012})}\BibitemShut
  {NoStop}%
\bibitem [{\citenamefont {Liu}\ \emph {et~al.}(2012)\citenamefont {Liu},
  \citenamefont {Pai}, \citenamefont {Li}, \citenamefont {Tseng}, \citenamefont
  {Ralph},\ and\ \citenamefont {Buhrman}}]{Liu:2012dga}%
  \BibitemOpen
  \bibfield  {author} {\bibinfo {author} {\bibfnamefont {L.}~\bibnamefont
  {Liu}}, \bibinfo {author} {\bibfnamefont {Chi-Feng}\ \bibnamefont {Pai}},
  \bibinfo {author} {\bibfnamefont {Y}~\bibnamefont {Li}}, \bibinfo {author}
  {\bibfnamefont {H~W}\ \bibnamefont {Tseng}}, \bibinfo {author} {\bibfnamefont
  {D.~C.}\ \bibnamefont {Ralph}}, \ and\ \bibinfo {author} {\bibfnamefont
  {R.~A.}\ \bibnamefont {Buhrman}},\ }\bibfield  {title} {\enquote {\bibinfo
  {title} {{Spin-Torque Switching with the Giant Spin Hall Effect of
  Tantalum}},}\ }\href@noop {} {\bibfield  {journal} {\bibinfo  {journal}
  {Science}\ }\textbf {\bibinfo {volume} {336}},\ \bibinfo {pages} {555--558}
  (\bibinfo {year} {2012})}\BibitemShut {NoStop}%
\bibitem [{\citenamefont {Garello}\ \emph {et~al.}(2014)\citenamefont
  {Garello}, \citenamefont {Avci}, \citenamefont {Miron}, \citenamefont
  {Baumgartner}, \citenamefont {Ghosh}, \citenamefont {Auffret}, \citenamefont
  {Boulle}, \citenamefont {Gaudin},\ and\ \citenamefont
  {Gambardella}}]{Garello:2014bi}%
  \BibitemOpen
  \bibfield  {author} {\bibinfo {author} {\bibfnamefont {K.}~\bibnamefont
  {Garello}}, \bibinfo {author} {\bibfnamefont {C.~O.}\ \bibnamefont {Avci}},
  \bibinfo {author} {\bibfnamefont {I.~M.}\ \bibnamefont {Miron}}, \bibinfo
  {author} {\bibfnamefont {Manuel}\ \bibnamefont {Baumgartner}}, \bibinfo
  {author} {\bibfnamefont {A.}~\bibnamefont {Ghosh}}, \bibinfo {author}
  {\bibfnamefont {S.}~\bibnamefont {Auffret}}, \bibinfo {author} {\bibfnamefont
  {O.}~\bibnamefont {Boulle}}, \bibinfo {author} {\bibfnamefont
  {G.}~\bibnamefont {Gaudin}}, \ and\ \bibinfo {author} {\bibfnamefont
  {P.}~\bibnamefont {Gambardella}},\ }\bibfield  {title} {\enquote {\bibinfo
  {title} {{Ultrafast magnetization switching by spin-orbit torques}},}\
  }\href@noop {} {\bibfield  {journal} {\bibinfo  {journal} {Appl. Phys.
  Lett.}\ }\textbf {\bibinfo {volume} {105}},\ \bibinfo {pages} {212402}
  (\bibinfo {year} {2014})}\BibitemShut {NoStop}%
\bibitem [{\citenamefont {Ciccarelli}\ \emph {et~al.}(2016)\citenamefont
  {Ciccarelli}, \citenamefont {Anderson}, \citenamefont {Tshitoyan},
  \citenamefont {Ferguson}, \citenamefont {Gerhard}, \citenamefont {Gould},
  \citenamefont {Molenkamp}, \citenamefont {Gayles}, \citenamefont {{\v
  Z}elezn{\'{y}}}, \citenamefont {{\v{S}}mejkal}, \citenamefont {Yuan},
  \citenamefont {Sinova}, \citenamefont {Freimuth},\ and\ \citenamefont
  {Jungwirth}}]{Ciccarelli:2016ju}%
  \BibitemOpen
  \bibfield  {author} {\bibinfo {author} {\bibfnamefont {C}~\bibnamefont
  {Ciccarelli}}, \bibinfo {author} {\bibfnamefont {L}~\bibnamefont {Anderson}},
  \bibinfo {author} {\bibfnamefont {V}~\bibnamefont {Tshitoyan}}, \bibinfo
  {author} {\bibfnamefont {A.~J.}\ \bibnamefont {Ferguson}}, \bibinfo {author}
  {\bibfnamefont {F}~\bibnamefont {Gerhard}}, \bibinfo {author} {\bibfnamefont
  {C}~\bibnamefont {Gould}}, \bibinfo {author} {\bibfnamefont {L~W}\
  \bibnamefont {Molenkamp}}, \bibinfo {author} {\bibfnamefont {J}~\bibnamefont
  {Gayles}}, \bibinfo {author} {\bibfnamefont {J.}~\bibnamefont {{\v
  Z}elezn{\'{y}}}}, \bibinfo {author} {\bibfnamefont {L}~\bibnamefont
  {{\v{S}}mejkal}}, \bibinfo {author} {\bibfnamefont {Z}~\bibnamefont {Yuan}},
  \bibinfo {author} {\bibfnamefont {J.}~\bibnamefont {Sinova}}, \bibinfo
  {author} {\bibfnamefont {F.}~\bibnamefont {Freimuth}}, \ and\ \bibinfo
  {author} {\bibfnamefont {T.}~\bibnamefont {Jungwirth}},\ }\bibfield  {title}
  {\enquote {\bibinfo {title} {{Room-temperature spin{\textendash}orbit torque
  in NiMnSb}},}\ }\href@noop {} {\bibfield  {journal} {\bibinfo  {journal}
  {Nature Phys.}\ } (\bibinfo {year} {2016})}\BibitemShut {NoStop}%
\bibitem [{\citenamefont {Fukami}\ \emph {et~al.}(2016)\citenamefont {Fukami},
  \citenamefont {Zhang}, \citenamefont {DuttaGupta}, \citenamefont {Kurenkov},\
  and\ \citenamefont {Ohno}}]{Fukami:2016ca}%
  \BibitemOpen
  \bibfield  {author} {\bibinfo {author} {\bibfnamefont {Shunsuke}\
  \bibnamefont {Fukami}}, \bibinfo {author} {\bibfnamefont {Chaoliang}\
  \bibnamefont {Zhang}}, \bibinfo {author} {\bibfnamefont {Samik}\ \bibnamefont
  {DuttaGupta}}, \bibinfo {author} {\bibfnamefont {Aleksandr}\ \bibnamefont
  {Kurenkov}}, \ and\ \bibinfo {author} {\bibfnamefont {Hideo}\ \bibnamefont
  {Ohno}},\ }\bibfield  {title} {\enquote {\bibinfo {title} {{Magnetization
  switching by spin{\textendash}orbit torque in an
  antiferromagnet{\textendash}ferromagnet bilayer system}},}\ }\href@noop {}
  {\bibfield  {journal} {\bibinfo  {journal} {Nat. Mater.}\ }\textbf {\bibinfo
  {volume} {15}},\ \bibinfo {pages} {535--541} (\bibinfo {year}
  {2016})}\BibitemShut {NoStop}%
\bibitem [{\citenamefont {Miron}\ \emph {et~al.}(2010)\citenamefont {Miron},
  \citenamefont {Gaudin}, \citenamefont {Auffret}, \citenamefont {Rodmacq},
  \citenamefont {Schuhl}, \citenamefont {Pizzini}, \citenamefont {Vogel},\ and\
  \citenamefont {Gambardella}}]{MihaiMiron:gv}%
  \BibitemOpen
  \bibfield  {author} {\bibinfo {author} {\bibfnamefont {I.~M.}\ \bibnamefont
  {Miron}}, \bibinfo {author} {\bibfnamefont {G.}~\bibnamefont {Gaudin}},
  \bibinfo {author} {\bibfnamefont {S.}~\bibnamefont {Auffret}}, \bibinfo
  {author} {\bibfnamefont {B}~\bibnamefont {Rodmacq}}, \bibinfo {author}
  {\bibfnamefont {A}~\bibnamefont {Schuhl}}, \bibinfo {author} {\bibfnamefont
  {Stefania}\ \bibnamefont {Pizzini}}, \bibinfo {author} {\bibfnamefont {Jan}\
  \bibnamefont {Vogel}}, \ and\ \bibinfo {author} {\bibfnamefont {Pietro}\
  \bibnamefont {Gambardella}},\ }\bibfield  {title} {\enquote {\bibinfo {title}
  {{Current-driven spin torque induced by the Rashba effect in a ferromagnetic
  metal layer}},}\ }\href@noop {} {\bibfield  {journal} {\bibinfo  {journal}
  {Nat. Mater.}\ }\textbf {\bibinfo {volume} {9}},\ \bibinfo {pages} {230--234}
  (\bibinfo {year} {2010})}\BibitemShut {NoStop}%
\bibitem [{\citenamefont {Kim}\ \emph {et~al.}(2016)\citenamefont {Kim},
  \citenamefont {Sheng}, \citenamefont {Takahashi}, \citenamefont {Mitani},\
  and\ \citenamefont {Hayashi}}]{Kim:2016di}%
  \BibitemOpen
  \bibfield  {author} {\bibinfo {author} {\bibfnamefont {Junyeon}\ \bibnamefont
  {Kim}}, \bibinfo {author} {\bibfnamefont {Peng}\ \bibnamefont {Sheng}},
  \bibinfo {author} {\bibfnamefont {S.}~\bibnamefont {Takahashi}}, \bibinfo
  {author} {\bibfnamefont {S.}~\bibnamefont {Mitani}}, \ and\ \bibinfo {author}
  {\bibfnamefont {Masamitsu}\ \bibnamefont {Hayashi}},\ }\bibfield  {title}
  {\enquote {\bibinfo {title} {{Spin Hall Magnetoresistance in Metallic
  Bilayers}},}\ }\href@noop {} {\bibfield  {journal} {\bibinfo  {journal}
  {Phys. Rev. Lett.}\ }\textbf {\bibinfo {volume} {116}},\ \bibinfo {pages}
  {097201} (\bibinfo {year} {2016})}\BibitemShut {NoStop}%
\bibitem [{\citenamefont {Kurebayashi}\ \emph {et~al.}(2014)\citenamefont
  {Kurebayashi}, \citenamefont {Sinova}, \citenamefont {Fang}, \citenamefont
  {Irvine}, \citenamefont {Skinner}, \citenamefont {Wunderlich}, \citenamefont
  {Nov{\'a}k}, \citenamefont {Campion}, \citenamefont {Gallagher},
  \citenamefont {Vehstedt}, \citenamefont {Z{\^a}rbo}, \citenamefont
  {V{\'{y}}born{\'{y}}}, \citenamefont {Ferguson},\ and\ \citenamefont
  {Jungwirth}}]{Kurebayashi:2014br}%
  \BibitemOpen
  \bibfield  {author} {\bibinfo {author} {\bibfnamefont {H.}~\bibnamefont
  {Kurebayashi}}, \bibinfo {author} {\bibfnamefont {J.}~\bibnamefont {Sinova}},
  \bibinfo {author} {\bibfnamefont {D.}~\bibnamefont {Fang}}, \bibinfo {author}
  {\bibfnamefont {A~C}\ \bibnamefont {Irvine}}, \bibinfo {author}
  {\bibfnamefont {T~D}\ \bibnamefont {Skinner}}, \bibinfo {author}
  {\bibfnamefont {J.}~\bibnamefont {Wunderlich}}, \bibinfo {author}
  {\bibfnamefont {V.}~\bibnamefont {Nov{\'a}k}}, \bibinfo {author}
  {\bibfnamefont {R.~P.}\ \bibnamefont {Campion}}, \bibinfo {author}
  {\bibfnamefont {B.~L.}\ \bibnamefont {Gallagher}}, \bibinfo {author}
  {\bibfnamefont {E~K}\ \bibnamefont {Vehstedt}}, \bibinfo {author}
  {\bibfnamefont {L.~P.}\ \bibnamefont {Z{\^a}rbo}}, \bibinfo {author}
  {\bibfnamefont {K.}~\bibnamefont {V{\'{y}}born{\'{y}}}}, \bibinfo {author}
  {\bibfnamefont {A.~J.}\ \bibnamefont {Ferguson}}, \ and\ \bibinfo {author}
  {\bibfnamefont {T.}~\bibnamefont {Jungwirth}},\ }\bibfield  {title} {\enquote
  {\bibinfo {title} {{An antidamping spin-orbit torque originating from the
  Berry curvature}},}\ }\href@noop {} {\bibfield  {journal} {\bibinfo
  {journal} {Nature Nanotechnol.}\ }\textbf {\bibinfo {volume} {9}},\ \bibinfo
  {pages} {211--217} (\bibinfo {year} {2014})}\BibitemShut {NoStop}%
\bibitem [{\citenamefont {Avci}\ \emph {et~al.}(2014)\citenamefont {Avci},
  \citenamefont {Garello}, \citenamefont {Nistor}, \citenamefont {Godey},
  \citenamefont {Ballesteros}, \citenamefont {Mugarza}, \citenamefont {Barla},
  \citenamefont {Valvidares}, \citenamefont {Pellegrin}, \citenamefont {Ghosh},
  \citenamefont {Miron}, \citenamefont {Boulle}, \citenamefont {Auffret},
  \citenamefont {Gaudin},\ and\ \citenamefont {Gambardella}}]{Avci:2014fj}%
  \BibitemOpen
  \bibfield  {author} {\bibinfo {author} {\bibfnamefont {C.~O.}\ \bibnamefont
  {Avci}}, \bibinfo {author} {\bibfnamefont {K.}~\bibnamefont {Garello}},
  \bibinfo {author} {\bibfnamefont {Corneliu}\ \bibnamefont {Nistor}}, \bibinfo
  {author} {\bibfnamefont {Sylvie}\ \bibnamefont {Godey}}, \bibinfo {author}
  {\bibfnamefont {Bel{\'e}n}\ \bibnamefont {Ballesteros}}, \bibinfo {author}
  {\bibfnamefont {Aitor}\ \bibnamefont {Mugarza}}, \bibinfo {author}
  {\bibfnamefont {Alessandro}\ \bibnamefont {Barla}}, \bibinfo {author}
  {\bibfnamefont {Manuel}\ \bibnamefont {Valvidares}}, \bibinfo {author}
  {\bibfnamefont {Eric}\ \bibnamefont {Pellegrin}}, \bibinfo {author}
  {\bibfnamefont {A.}~\bibnamefont {Ghosh}}, \bibinfo {author} {\bibfnamefont
  {I.~M.}\ \bibnamefont {Miron}}, \bibinfo {author} {\bibfnamefont
  {O.}~\bibnamefont {Boulle}}, \bibinfo {author} {\bibfnamefont
  {S.}~\bibnamefont {Auffret}}, \bibinfo {author} {\bibfnamefont
  {G.}~\bibnamefont {Gaudin}}, \ and\ \bibinfo {author} {\bibfnamefont
  {P.}~\bibnamefont {Gambardella}},\ }\bibfield  {title} {\enquote {\bibinfo
  {title} {{Fieldlike and antidamping spin-orbit torques in as-grown and
  annealed Ta/CoFeB/MgO layers}},}\ }\href@noop {} {\bibfield  {journal}
  {\bibinfo  {journal} {Phys. Rev. B}\ }\textbf {\bibinfo {volume} {89}},\
  \bibinfo {pages} {214419} (\bibinfo {year} {2014})}\BibitemShut {NoStop}%
\bibitem [{\citenamefont {Hayashi}\ \emph {et~al.}(2014)\citenamefont
  {Hayashi}, \citenamefont {Kim}, \citenamefont {Yamanouchi},\ and\
  \citenamefont {Ohno}}]{Hayashi:2014bd}%
  \BibitemOpen
  \bibfield  {author} {\bibinfo {author} {\bibfnamefont {Masamitsu}\
  \bibnamefont {Hayashi}}, \bibinfo {author} {\bibfnamefont {Junyeon}\
  \bibnamefont {Kim}}, \bibinfo {author} {\bibfnamefont {Michihiko}\
  \bibnamefont {Yamanouchi}}, \ and\ \bibinfo {author} {\bibfnamefont {Hideo}\
  \bibnamefont {Ohno}},\ }\bibfield  {title} {\enquote {\bibinfo {title}
  {{Quantitative characterization of the spin-orbit torque using harmonic Hall
  voltage measurements}},}\ }\href@noop {} {\bibfield  {journal} {\bibinfo
  {journal} {Phys. Rev. B}\ }\textbf {\bibinfo {volume} {89}},\ \bibinfo
  {pages} {144425} (\bibinfo {year} {2014})}\BibitemShut {NoStop}%
\bibitem [{\citenamefont {McGuire}\ and\ \citenamefont
  {Potter}(1975)}]{McGuire:1975jk}%
  \BibitemOpen
  \bibfield  {author} {\bibinfo {author} {\bibfnamefont {T}~\bibnamefont
  {McGuire}}\ and\ \bibinfo {author} {\bibfnamefont {R}~\bibnamefont
  {Potter}},\ }\bibfield  {title} {\enquote {\bibinfo {title} {{Anisotropic
  magnetoresistance in ferromagnetic 3d alloys}},}\ }\href@noop {} {\bibfield
  {journal} {\bibinfo  {journal} {Magnetics, IEEE Transactions on}\ }\textbf
  {\bibinfo {volume} {11}},\ \bibinfo {pages} {1018--1038} (\bibinfo {year}
  {1975})}\BibitemShut {NoStop}%
\bibitem [{\citenamefont {Kupferschmidt}\ \emph {et~al.}(2006)\citenamefont
  {Kupferschmidt}, \citenamefont {Adam},\ and\ \citenamefont
  {Brouwer}}]{Kupferschmidt:2006ho}%
  \BibitemOpen
  \bibfield  {author} {\bibinfo {author} {\bibfnamefont {Joern~N}\ \bibnamefont
  {Kupferschmidt}}, \bibinfo {author} {\bibfnamefont {Shaffique}\ \bibnamefont
  {Adam}}, \ and\ \bibinfo {author} {\bibfnamefont {Piet~W}\ \bibnamefont
  {Brouwer}},\ }\bibfield  {title} {\enquote {\bibinfo {title} {{Theory of the
  spin-torque-driven ferromagnetic resonance in a
  ferromagnet/normal-metal/ferromagnet structure}},}\ }\href@noop {} {\bibfield
   {journal} {\bibinfo  {journal} {Phys. Rev. B}\ }\textbf {\bibinfo {volume}
  {74}},\ \bibinfo {pages} {134416} (\bibinfo {year} {2006})}\BibitemShut
  {NoStop}%
\bibitem [{\citenamefont {Kovalev}\ \emph {et~al.}(2007)\citenamefont
  {Kovalev}, \citenamefont {Bauer},\ and\ \citenamefont
  {Brataas}}]{Kovalev:2007hk}%
  \BibitemOpen
  \bibfield  {author} {\bibinfo {author} {\bibfnamefont {Alexey~A}\
  \bibnamefont {Kovalev}}, \bibinfo {author} {\bibfnamefont {G.~E.~W.}\
  \bibnamefont {Bauer}}, \ and\ \bibinfo {author} {\bibfnamefont {Arne}\
  \bibnamefont {Brataas}},\ }\bibfield  {title} {\enquote {\bibinfo {title}
  {{Current-driven ferromagnetic resonance, mechanical torques, and rotary
  motion in magnetic nanostructures}},}\ }\href@noop {} {\bibfield  {journal}
  {\bibinfo  {journal} {Phys. Rev. B}\ }\textbf {\bibinfo {volume} {75}},\
  \bibinfo {pages} {014430} (\bibinfo {year} {2007})}\BibitemShut {NoStop}%
\bibitem [{\citenamefont {Tserkovnyak}\ and\ \citenamefont
  {Bender}(2014)}]{Tserkovnyak:2014jd}%
  \BibitemOpen
  \bibfield  {author} {\bibinfo {author} {\bibfnamefont {Yaroslav}\
  \bibnamefont {Tserkovnyak}}\ and\ \bibinfo {author} {\bibfnamefont {Scott~A}\
  \bibnamefont {Bender}},\ }\bibfield  {title} {\enquote {\bibinfo {title}
  {{Spin Hall phenomenology of magnetic dynamics}},}\ }\href@noop {} {\bibfield
   {journal} {\bibinfo  {journal} {Phys. Rev. B}\ }\textbf {\bibinfo {volume}
  {90}},\ \bibinfo {pages} {014428} (\bibinfo {year} {2014})}\BibitemShut
  {NoStop}%
\bibitem [{\citenamefont {Ciccarelli}\ \emph {et~al.}(2014)\citenamefont
  {Ciccarelli}, \citenamefont {Hals}, \citenamefont {Irvine}, \citenamefont
  {Novak}, \citenamefont {Tserkovnyak}, \citenamefont {Kurebayashi},
  \citenamefont {Brataas},\ and\ \citenamefont {Ferguson}}]{Ciccarelli:2014bl}%
  \BibitemOpen
  \bibfield  {author} {\bibinfo {author} {\bibfnamefont {Chiara}\ \bibnamefont
  {Ciccarelli}}, \bibinfo {author} {\bibfnamefont {Kjetil M~D}\ \bibnamefont
  {Hals}}, \bibinfo {author} {\bibfnamefont {Andrew}\ \bibnamefont {Irvine}},
  \bibinfo {author} {\bibfnamefont {V.}~\bibnamefont {Novak}}, \bibinfo
  {author} {\bibfnamefont {Yaroslav}\ \bibnamefont {Tserkovnyak}}, \bibinfo
  {author} {\bibfnamefont {H.}~\bibnamefont {Kurebayashi}}, \bibinfo {author}
  {\bibfnamefont {Arne}\ \bibnamefont {Brataas}}, \ and\ \bibinfo {author}
  {\bibfnamefont {A.}~\bibnamefont {Ferguson}},\ }\bibfield  {title} {\enquote
  {\bibinfo {title} {{Magnonic charge pumping via spin{\textendash}orbit
  coupling}},}\ }\href@noop {} {\bibfield  {journal} {\bibinfo  {journal}
  {Nature Nanotechnol.}\ }\textbf {\bibinfo {volume} {10}},\ \bibinfo {pages}
  {50--54} (\bibinfo {year} {2014})}\BibitemShut {NoStop}%
\bibitem [{\citenamefont {Nagaosa}\ \emph {et~al.}(2010)\citenamefont
  {Nagaosa}, \citenamefont {Onoda}, \citenamefont {MacDonald},\ and\
  \citenamefont {Ong}}]{Nagaosa:2010js}%
  \BibitemOpen
  \bibfield  {author} {\bibinfo {author} {\bibfnamefont {N.}~\bibnamefont
  {Nagaosa}}, \bibinfo {author} {\bibfnamefont {Shigeki}\ \bibnamefont
  {Onoda}}, \bibinfo {author} {\bibfnamefont {A~H}\ \bibnamefont {MacDonald}},
  \ and\ \bibinfo {author} {\bibfnamefont {N~P}\ \bibnamefont {Ong}},\
  }\bibfield  {title} {\enquote {\bibinfo {title} {{Anomalous Hall effect}},}\
  }\href@noop {} {\bibfield  {journal} {\bibinfo  {journal} {Rev. Mod. Phys.}\
  }\textbf {\bibinfo {volume} {82}},\ \bibinfo {pages} {1539--1592} (\bibinfo
  {year} {2010})}\BibitemShut {NoStop}%
\bibitem [{\citenamefont {Seemann}\ \emph {et~al.}(2011)\citenamefont
  {Seemann}, \citenamefont {Freimuth}, \citenamefont {Zhang}, \citenamefont
  {Bl{\"u}gel}, \citenamefont {Mokrousov}, \citenamefont {B{\"u}rgler},\ and\
  \citenamefont {Schneider}}]{Seemann:2011fu}%
  \BibitemOpen
  \bibfield  {author} {\bibinfo {author} {\bibfnamefont {K~M}\ \bibnamefont
  {Seemann}}, \bibinfo {author} {\bibfnamefont {F.}~\bibnamefont {Freimuth}},
  \bibinfo {author} {\bibfnamefont {H.}~\bibnamefont {Zhang}}, \bibinfo
  {author} {\bibfnamefont {S.}~\bibnamefont {Bl{\"u}gel}}, \bibinfo {author}
  {\bibfnamefont {Y.}~\bibnamefont {Mokrousov}}, \bibinfo {author}
  {\bibfnamefont {D~E}\ \bibnamefont {B{\"u}rgler}}, \ and\ \bibinfo {author}
  {\bibfnamefont {C~M}\ \bibnamefont {Schneider}},\ }\bibfield  {title}
  {\enquote {\bibinfo {title} {{Origin of the Planar Hall Effect in
  Nanocrystalline Co60Fe20B20}},}\ }\href@noop {} {\bibfield  {journal}
  {\bibinfo  {journal} {Phys. Rev. Lett.}\ }\textbf {\bibinfo {volume} {107}},\
  \bibinfo {pages} {086603} (\bibinfo {year} {2011})}\BibitemShut {NoStop}%
\bibitem [{\citenamefont {Bouaziz}\ \emph {et~al.}(2016)\citenamefont
  {Bouaziz}, \citenamefont {Lounis}, \citenamefont {Bl{\"u}gel},\ and\
  \citenamefont {Ishida}}]{Bouaziz:2016fb}%
  \BibitemOpen
  \bibfield  {author} {\bibinfo {author} {\bibfnamefont {Juba}\ \bibnamefont
  {Bouaziz}}, \bibinfo {author} {\bibfnamefont {Samir}\ \bibnamefont {Lounis}},
  \bibinfo {author} {\bibfnamefont {S.}~\bibnamefont {Bl{\"u}gel}}, \ and\
  \bibinfo {author} {\bibfnamefont {Hiroshi}\ \bibnamefont {Ishida}},\
  }\bibfield  {title} {\enquote {\bibinfo {title} {{Microscopic theory of the
  residual surface resistivity of Rashba electrons}},}\ }\href@noop {}
  {\bibfield  {journal} {\bibinfo  {journal} {Phys. Rev. B}\ }\textbf {\bibinfo
  {volume} {94}},\ \bibinfo {pages} {045433} (\bibinfo {year}
  {2016})}\BibitemShut {NoStop}%
\bibitem [{\citenamefont {Tserkovnyak}\ \emph {et~al.}(2002)\citenamefont
  {Tserkovnyak}, \citenamefont {Brataas},\ and\ \citenamefont
  {Bauer}}]{Tserkovnyak:2002ju}%
  \BibitemOpen
  \bibfield  {author} {\bibinfo {author} {\bibfnamefont {Yaroslav}\
  \bibnamefont {Tserkovnyak}}, \bibinfo {author} {\bibfnamefont {Arne}\
  \bibnamefont {Brataas}}, \ and\ \bibinfo {author} {\bibfnamefont {G.~E.~W.}\
  \bibnamefont {Bauer}},\ }\bibfield  {title} {\enquote {\bibinfo {title}
  {{Enhanced Gilbert Damping in Thin Ferromagnetic Films}},}\ }\href@noop {}
  {\bibfield  {journal} {\bibinfo  {journal} {Phys. Rev. Lett.}\ }\textbf
  {\bibinfo {volume} {88}},\ \bibinfo {pages} {117601} (\bibinfo {year}
  {2002})}\BibitemShut {NoStop}%
\bibitem [{\citenamefont {Hoffmann}(2013)}]{Hoffmann:el}%
  \BibitemOpen
  \bibfield  {author} {\bibinfo {author} {\bibfnamefont {A.}~\bibnamefont
  {Hoffmann}},\ }\bibfield  {title} {\enquote {\bibinfo {title} {{Spin Hall
  Effects in Metals}},}\ }\href@noop {} {\bibfield  {journal} {\bibinfo
  {journal} {Magnetics, IEEE Transactions on}\ }\textbf {\bibinfo {volume}
  {49}},\ \bibinfo {pages} {5172--5193} (\bibinfo {year} {2013})}\BibitemShut
  {NoStop}%
\bibitem [{\citenamefont {Smit}\ and\ \citenamefont
  {Beljers}(1955)}]{Smit:1955wl}%
  \BibitemOpen
  \bibfield  {author} {\bibinfo {author} {\bibfnamefont {J}~\bibnamefont
  {Smit}}\ and\ \bibinfo {author} {\bibfnamefont {H~G}\ \bibnamefont
  {Beljers}},\ }\bibfield  {title} {\enquote {\bibinfo {title} {{Ferromagnetic
  resonance absorption in BaFe$_{12}$O$_{19}$, a highly anisotropic
  crystal}},}\ }\href@noop {} {\bibfield  {journal} {\bibinfo  {journal}
  {Philips Res. Rep.}\ }\textbf {\bibinfo {volume} {10}},\ \bibinfo {pages}
  {113--130} (\bibinfo {year} {1955})}\BibitemShut {NoStop}%
\bibitem [{\citenamefont {Farle}(1998)}]{Farle:1998gz}%
  \BibitemOpen
  \bibfield  {author} {\bibinfo {author} {\bibfnamefont {Michael}\ \bibnamefont
  {Farle}},\ }\bibfield  {title} {\enquote {\bibinfo {title} {{Ferromagnetic
  resonance of ultrathin metallic layers}},}\ }\href@noop {} {\bibfield
  {journal} {\bibinfo  {journal} {Reports on Progress in Physics}\ }\textbf
  {\bibinfo {volume} {61}},\ \bibinfo {pages} {755--826} (\bibinfo {year}
  {1998})}\BibitemShut {NoStop}%
\bibitem [{\citenamefont {dos Santos~Dias}\ \emph {et~al.}(2015)\citenamefont
  {dos Santos~Dias}, \citenamefont {Schweflinghaus}, \citenamefont
  {Bl{\"u}gel},\ and\ \citenamefont {Lounis}}]{dosSantosDias:2015bh}%
  \BibitemOpen
  \bibfield  {author} {\bibinfo {author} {\bibfnamefont {M.}~\bibnamefont {dos
  Santos~Dias}}, \bibinfo {author} {\bibfnamefont {B}~\bibnamefont
  {Schweflinghaus}}, \bibinfo {author} {\bibfnamefont {S.}~\bibnamefont
  {Bl{\"u}gel}}, \ and\ \bibinfo {author} {\bibfnamefont {S.}~\bibnamefont
  {Lounis}},\ }\bibfield  {title} {\enquote {\bibinfo {title} {{Relativistic
  dynamical spin excitations of magnetic adatoms}},}\ }\href@noop {} {\bibfield
   {journal} {\bibinfo  {journal} {Phys. Rev. B}\ }\textbf {\bibinfo {volume}
  {91}},\ \bibinfo {pages} {075405} (\bibinfo {year} {2015})}\BibitemShut
  {NoStop}%
\bibitem [{\citenamefont {Lehnert}\ \emph {et~al.}(2010)\citenamefont
  {Lehnert}, \citenamefont {Dennler}, \citenamefont {B{\l}o{\'{n}}ski},
  \citenamefont {Rusponi}, \citenamefont {Etzkorn}, \citenamefont {Moulas},
  \citenamefont {Bencok}, \citenamefont {Gambardella}, \citenamefont {Brune},\
  and\ \citenamefont {Hafner}}]{Lehnert:2010fu}%
  \BibitemOpen
  \bibfield  {author} {\bibinfo {author} {\bibfnamefont {Anne}\ \bibnamefont
  {Lehnert}}, \bibinfo {author} {\bibfnamefont {Samuel}\ \bibnamefont
  {Dennler}}, \bibinfo {author} {\bibfnamefont {Piotr}\ \bibnamefont
  {B{\l}o{\'{n}}ski}}, \bibinfo {author} {\bibfnamefont {Stefano}\ \bibnamefont
  {Rusponi}}, \bibinfo {author} {\bibfnamefont {Markus}\ \bibnamefont
  {Etzkorn}}, \bibinfo {author} {\bibfnamefont {G{\'e}raud}\ \bibnamefont
  {Moulas}}, \bibinfo {author} {\bibfnamefont {Peter}\ \bibnamefont {Bencok}},
  \bibinfo {author} {\bibfnamefont {P.}~\bibnamefont {Gambardella}}, \bibinfo
  {author} {\bibfnamefont {Harald}\ \bibnamefont {Brune}}, \ and\ \bibinfo
  {author} {\bibfnamefont {J{\"u}rgen}\ \bibnamefont {Hafner}},\ }\bibfield
  {title} {\enquote {\bibinfo {title} {{Magnetic anisotropy of Fe and Co
  ultrathin films deposited on Rh(111) and Pt(111) substrates: An experimental
  and first-principles investigation}},}\ }\href@noop {} {\bibfield  {journal}
  {\bibinfo  {journal} {Phys. Rev. B}\ }\textbf {\bibinfo {volume} {82}},\
  \bibinfo {pages} {094409} (\bibinfo {year} {2010})}\BibitemShut {NoStop}%
\bibitem [{\citenamefont {{\v{S}}ipr}\ \emph {et~al.}(2010)\citenamefont
  {{\v{S}}ipr}, \citenamefont {Bornemann}, \citenamefont {Min{\'a}r},\ and\
  \citenamefont {Ebert}}]{Sipr:2010di}%
  \BibitemOpen
  \bibfield  {author} {\bibinfo {author} {\bibfnamefont {O}~\bibnamefont
  {{\v{S}}ipr}}, \bibinfo {author} {\bibfnamefont {S.}~\bibnamefont
  {Bornemann}}, \bibinfo {author} {\bibfnamefont {J}~\bibnamefont {Min{\'a}r}},
  \ and\ \bibinfo {author} {\bibfnamefont {H}~\bibnamefont {Ebert}},\
  }\bibfield  {title} {\enquote {\bibinfo {title} {{Magnetic anisotropy of Fe
  and Co adatoms and monolayers: Need for a proper treatment of the
  substrate}},}\ }\href@noop {} {\bibfield  {journal} {\bibinfo  {journal}
  {Phys. Rev. B}\ }\textbf {\bibinfo {volume} {82}},\ \bibinfo {pages} {174414}
  (\bibinfo {year} {2010})}\BibitemShut {NoStop}%
\bibitem [{\citenamefont {Rojas~S{\'a}nchez}\ \emph {et~al.}(2014)\citenamefont
  {Rojas~S{\'a}nchez}, \citenamefont {Reyren}, \citenamefont {Laczkowski},
  \citenamefont {Savero}, \citenamefont {Attan{\'e}}, \citenamefont {Deranlot},
  \citenamefont {Jamet}, \citenamefont {George}, \citenamefont {Vila},\ and\
  \citenamefont {Jaffr{\`e}s}}]{RojasSanchez:2014ih}%
  \BibitemOpen
  \bibfield  {author} {\bibinfo {author} {\bibfnamefont {J.~C.}\ \bibnamefont
  {Rojas~S{\'a}nchez}}, \bibinfo {author} {\bibfnamefont {N}~\bibnamefont
  {Reyren}}, \bibinfo {author} {\bibfnamefont {P}~\bibnamefont {Laczkowski}},
  \bibinfo {author} {\bibfnamefont {W}~\bibnamefont {Savero}}, \bibinfo
  {author} {\bibfnamefont {J.~P.}\ \bibnamefont {Attan{\'e}}}, \bibinfo
  {author} {\bibfnamefont {C}~\bibnamefont {Deranlot}}, \bibinfo {author}
  {\bibfnamefont {M}~\bibnamefont {Jamet}}, \bibinfo {author} {\bibfnamefont
  {J~M}\ \bibnamefont {George}}, \bibinfo {author} {\bibfnamefont
  {L.}~\bibnamefont {Vila}}, \ and\ \bibinfo {author} {\bibfnamefont
  {H}~\bibnamefont {Jaffr{\`e}s}},\ }\bibfield  {title} {\enquote {\bibinfo
  {title} {{Spin Pumping and Inverse Spin Hall Effect in Platinum: The
  Essential Role of Spin-Memory Loss at Metallic Interfaces}},}\ }\href@noop {}
  {\bibfield  {journal} {\bibinfo  {journal} {Phys. Rev. Lett.}\ }\textbf
  {\bibinfo {volume} {112}},\ \bibinfo {pages} {106602} (\bibinfo {year}
  {2014})}\BibitemShut {NoStop}%
\bibitem [{\citenamefont {Weiler}\ \emph {et~al.}(2014)\citenamefont {Weiler},
  \citenamefont {Shaw}, \citenamefont {Nembach},\ and\ \citenamefont
  {Silva}}]{Weiler:2014bp}%
  \BibitemOpen
  \bibfield  {author} {\bibinfo {author} {\bibfnamefont {Mathias}\ \bibnamefont
  {Weiler}}, \bibinfo {author} {\bibfnamefont {Justin~M}\ \bibnamefont {Shaw}},
  \bibinfo {author} {\bibfnamefont {Hans~T}\ \bibnamefont {Nembach}}, \ and\
  \bibinfo {author} {\bibfnamefont {Thomas~J}\ \bibnamefont {Silva}},\
  }\bibfield  {title} {\enquote {\bibinfo {title} {{Phase-Sensitive Detection
  of Spin Pumping via the ac Inverse Spin Hall Effect}},}\ }\href@noop {}
  {\bibfield  {journal} {\bibinfo  {journal} {Phys. Rev. Lett.}\ }\textbf
  {\bibinfo {volume} {113}},\ \bibinfo {pages} {157204} (\bibinfo {year}
  {2014})}\BibitemShut {NoStop}%
\bibitem [{\citenamefont {Lau}\ \emph {et~al.}(2016)\citenamefont {Lau},
  \citenamefont {Betto}, \citenamefont {Rode}, \citenamefont {Coey},\ and\
  \citenamefont {Stamenov}}]{Lau:2016hz}%
  \BibitemOpen
  \bibfield  {author} {\bibinfo {author} {\bibfnamefont {Yong-Chang}\
  \bibnamefont {Lau}}, \bibinfo {author} {\bibfnamefont {Davide}\ \bibnamefont
  {Betto}}, \bibinfo {author} {\bibfnamefont {Karsten}\ \bibnamefont {Rode}},
  \bibinfo {author} {\bibfnamefont {J~M~D}\ \bibnamefont {Coey}}, \ and\
  \bibinfo {author} {\bibfnamefont {Plamen}\ \bibnamefont {Stamenov}},\
  }\bibfield  {title} {\enquote {\bibinfo {title} {{Spin{\textendash}orbit
  torque switching without an external field using interlayer exchange
  coupling}},}\ }\href@noop {} {\bibfield  {journal} {\bibinfo  {journal}
  {Nature Nanotechnol.}\ }\textbf {\bibinfo {volume} {11}},\ \bibinfo {pages}
  {758--762} (\bibinfo {year} {2016})}\BibitemShut {NoStop}%
\bibitem [{\citenamefont {Ghosh}\ \emph {et~al.}(2017)\citenamefont {Ghosh},
  \citenamefont {Garello}, \citenamefont {Avci}, \citenamefont {Gabureac},\
  and\ \citenamefont {Gambardella}}]{Ghosh:2017ib}%
  \BibitemOpen
  \bibfield  {author} {\bibinfo {author} {\bibfnamefont {A.}~\bibnamefont
  {Ghosh}}, \bibinfo {author} {\bibfnamefont {K.}~\bibnamefont {Garello}},
  \bibinfo {author} {\bibfnamefont {C.~O.}\ \bibnamefont {Avci}}, \bibinfo
  {author} {\bibfnamefont {M.}~\bibnamefont {Gabureac}}, \ and\ \bibinfo
  {author} {\bibfnamefont {P.}~\bibnamefont {Gambardella}},\ }\bibfield
  {title} {\enquote {\bibinfo {title} {{Interface-Enhanced Spin-Orbit Torques
  and Current-Induced Magnetization Switching of Pd/Co/AlO$_{x}$ Layers}},}\
  }\href@noop {} {\bibfield  {journal} {\bibinfo  {journal} {Phys. Rev.
  Applied}\ }\textbf {\bibinfo {volume} {7}},\ \bibinfo {pages} {014004}
  (\bibinfo {year} {2017})}\BibitemShut {NoStop}%
\bibitem [{\citenamefont {Kim}\ \emph {et~al.}(2012)\citenamefont {Kim},
  \citenamefont {Sinha}, \citenamefont {Hayashi}, \citenamefont {Yamanouchi},
  \citenamefont {Fukami}, \citenamefont {Suzuki}, \citenamefont {Mitani},\ and\
  \citenamefont {Ohno}}]{Kim:2012bk}%
  \BibitemOpen
  \bibfield  {author} {\bibinfo {author} {\bibfnamefont {Junyeon}\ \bibnamefont
  {Kim}}, \bibinfo {author} {\bibfnamefont {Jaivardhan}\ \bibnamefont {Sinha}},
  \bibinfo {author} {\bibfnamefont {Masamitsu}\ \bibnamefont {Hayashi}},
  \bibinfo {author} {\bibfnamefont {Michihiko}\ \bibnamefont {Yamanouchi}},
  \bibinfo {author} {\bibfnamefont {Shunsuke}\ \bibnamefont {Fukami}}, \bibinfo
  {author} {\bibfnamefont {Tetsuhiro}\ \bibnamefont {Suzuki}}, \bibinfo
  {author} {\bibfnamefont {S.}~\bibnamefont {Mitani}}, \ and\ \bibinfo {author}
  {\bibfnamefont {Hideo}\ \bibnamefont {Ohno}},\ }\bibfield  {title} {\enquote
  {\bibinfo {title} {{Layer thickness dependence of the current-induced
  effective field vector in Ta|CoFeB|MgO}},}\ }\href@noop {} {\bibfield
  {journal} {\bibinfo  {journal} {Nat. Mater.}\ }\textbf {\bibinfo {volume}
  {12}},\ \bibinfo {pages} {240--245} (\bibinfo {year} {2012})}\BibitemShut
  {NoStop}%
\bibitem [{\citenamefont {Zhang}\ \emph {et~al.}(2016)\citenamefont {Zhang},
  \citenamefont {Han}, \citenamefont {Yang}, \citenamefont {Sun}, \citenamefont
  {Zhang}, \citenamefont {Yan},\ and\ \citenamefont {Parkin}}]{Zhang:2016bf}%
  \BibitemOpen
  \bibfield  {author} {\bibinfo {author} {\bibfnamefont {W.}~\bibnamefont
  {Zhang}}, \bibinfo {author} {\bibfnamefont {W.}~\bibnamefont {Han}}, \bibinfo
  {author} {\bibfnamefont {S.-H.}\ \bibnamefont {Yang}}, \bibinfo {author}
  {\bibfnamefont {Yan}\ \bibnamefont {Sun}}, \bibinfo {author} {\bibfnamefont
  {Yang}\ \bibnamefont {Zhang}}, \bibinfo {author} {\bibfnamefont {Binghai}\
  \bibnamefont {Yan}}, \ and\ \bibinfo {author} {\bibfnamefont {S.~S.~P.}\
  \bibnamefont {Parkin}},\ }\bibfield  {title} {\enquote {\bibinfo {title}
  {{Giant facet-dependent spin-orbit torque and spin Hall conductivity in the
  triangular antiferromagnet IrMn3}},}\ }\href@noop {} {\bibfield  {journal}
  {\bibinfo  {journal} {Science Advances}\ }\textbf {\bibinfo {volume} {2}},\
  \bibinfo {pages} {e1600759--e1600759} (\bibinfo {year} {2016})}\BibitemShut
  {NoStop}%
\bibitem [{\citenamefont {Okamoto}\ \emph {et~al.}(2015)\citenamefont
  {Okamoto}, \citenamefont {Kikuchi}, \citenamefont {Furuta}, \citenamefont
  {Kitakami},\ and\ \citenamefont {Shimatsu}}]{Okamoto:2015gq}%
  \BibitemOpen
  \bibfield  {author} {\bibinfo {author} {\bibfnamefont {Satoshi}\ \bibnamefont
  {Okamoto}}, \bibinfo {author} {\bibfnamefont {Nobuaki}\ \bibnamefont
  {Kikuchi}}, \bibinfo {author} {\bibfnamefont {Masaki}\ \bibnamefont
  {Furuta}}, \bibinfo {author} {\bibfnamefont {Osamu}\ \bibnamefont
  {Kitakami}}, \ and\ \bibinfo {author} {\bibfnamefont {Takehito}\ \bibnamefont
  {Shimatsu}},\ }\bibfield  {title} {\enquote {\bibinfo {title} {{Microwave
  assisted magnetic recording technologies and related physics}},}\ }\href@noop
  {} {\bibfield  {journal} {\bibinfo  {journal} {J. Phys. D: Appl. Phys.}\
  }\textbf {\bibinfo {volume} {48}},\ \bibinfo {pages} {353001} (\bibinfo
  {year} {2015})}\BibitemShut {NoStop}%
\bibitem [{\citenamefont {Seifert}\ \emph {et~al.}(2016)\citenamefont
  {Seifert}, \citenamefont {Jaiswal}, \citenamefont {Martens}, \citenamefont
  {Hannegan}, \citenamefont {Braun}, \citenamefont {Maldonado}, \citenamefont
  {Freimuth}, \citenamefont {Kronenberg}, \citenamefont {Henrizi},
  \citenamefont {Radu}, \citenamefont {Beaurepaire}, \citenamefont {Mokrousov},
  \citenamefont {Oppeneer}, \citenamefont {Jourdan}, \citenamefont {Jakob},
  \citenamefont {Turchinovich}, \citenamefont {Hayden}, \citenamefont {Wolf},
  \citenamefont {M{\"u}nzenberg}, \citenamefont {Kl{\"a}ui},\ and\
  \citenamefont {Kampfrath}}]{SeifertT:2016kc}%
  \BibitemOpen
  \bibfield  {author} {\bibinfo {author} {\bibfnamefont {T}~\bibnamefont
  {Seifert}}, \bibinfo {author} {\bibfnamefont {S}~\bibnamefont {Jaiswal}},
  \bibinfo {author} {\bibfnamefont {U}~\bibnamefont {Martens}}, \bibinfo
  {author} {\bibfnamefont {J}~\bibnamefont {Hannegan}}, \bibinfo {author}
  {\bibfnamefont {L}~\bibnamefont {Braun}}, \bibinfo {author} {\bibfnamefont
  {P}~\bibnamefont {Maldonado}}, \bibinfo {author} {\bibfnamefont
  {F.}~\bibnamefont {Freimuth}}, \bibinfo {author} {\bibfnamefont
  {A}~\bibnamefont {Kronenberg}}, \bibinfo {author} {\bibfnamefont
  {J}~\bibnamefont {Henrizi}}, \bibinfo {author} {\bibfnamefont
  {I}~\bibnamefont {Radu}}, \bibinfo {author} {\bibfnamefont {E}~\bibnamefont
  {Beaurepaire}}, \bibinfo {author} {\bibfnamefont {Y.}~\bibnamefont
  {Mokrousov}}, \bibinfo {author} {\bibfnamefont {P~M}\ \bibnamefont
  {Oppeneer}}, \bibinfo {author} {\bibfnamefont {M}~\bibnamefont {Jourdan}},
  \bibinfo {author} {\bibfnamefont {G}~\bibnamefont {Jakob}}, \bibinfo {author}
  {\bibfnamefont {D}~\bibnamefont {Turchinovich}}, \bibinfo {author}
  {\bibfnamefont {L~M}\ \bibnamefont {Hayden}}, \bibinfo {author}
  {\bibfnamefont {M}~\bibnamefont {Wolf}}, \bibinfo {author} {\bibfnamefont
  {M}~\bibnamefont {M{\"u}nzenberg}}, \bibinfo {author} {\bibfnamefont
  {M}~\bibnamefont {Kl{\"a}ui}}, \ and\ \bibinfo {author} {\bibfnamefont
  {T}~\bibnamefont {Kampfrath}},\ }\bibfield  {title} {\enquote {\bibinfo
  {title} {{Efficient metallic spintronic emitters of ultrabroadband terahertz
  radiation}},}\ }\href@noop {} {\bibfield  {journal} {\bibinfo  {journal}
  {Nature Photon}\ }\textbf {\bibinfo {volume} {10}},\ \bibinfo {pages}
  {483--488} (\bibinfo {year} {2016})}\BibitemShut {NoStop}%
\bibitem [{\citenamefont {Andersen}\ and\ \citenamefont
  {Jepsen}(1984)}]{Andersen:1984ir}%
  \BibitemOpen
  \bibfield  {author} {\bibinfo {author} {\bibfnamefont {O~K}\ \bibnamefont
  {Andersen}}\ and\ \bibinfo {author} {\bibfnamefont {O}~\bibnamefont
  {Jepsen}},\ }\bibfield  {title} {\enquote {\bibinfo {title} {{Explicit,
  First-Principles Tight-Binding Theory}},}\ }\href@noop {} {\bibfield
  {journal} {\bibinfo  {journal} {Phys. Rev. Lett.}\ }\textbf {\bibinfo
  {volume} {53}},\ \bibinfo {pages} {2571--2574} (\bibinfo {year}
  {1984})}\BibitemShut {NoStop}%
\bibitem [{\citenamefont {Peduto}\ \emph {et~al.}(1991)\citenamefont {Peduto},
  \citenamefont {Frota-Pess{\^o}a},\ and\ \citenamefont
  {Methfessel}}]{Peduto:1991hn}%
  \BibitemOpen
  \bibfield  {author} {\bibinfo {author} {\bibfnamefont {Pascoal~R}\
  \bibnamefont {Peduto}}, \bibinfo {author} {\bibfnamefont {Sonia}\
  \bibnamefont {Frota-Pess{\^o}a}}, \ and\ \bibinfo {author} {\bibfnamefont
  {Michael~S}\ \bibnamefont {Methfessel}},\ }\bibfield  {title} {\enquote
  {\bibinfo {title} {{First-principles linear muffin-tin orbital atomic-sphere
  approximation calculations in real space}},}\ }\href@noop {} {\bibfield
  {journal} {\bibinfo  {journal} {Phys. Rev. B}\ }\textbf {\bibinfo {volume}
  {44}},\ \bibinfo {pages} {13283--13290} (\bibinfo {year} {1991})}\BibitemShut
  {NoStop}%
\bibitem [{\citenamefont {Frota-Pess{\^o}a}(1992)}]{FrotaPessoa:1992kt}%
  \BibitemOpen
  \bibfield  {author} {\bibinfo {author} {\bibfnamefont {Sonia}\ \bibnamefont
  {Frota-Pess{\^o}a}},\ }\bibfield  {title} {\enquote {\bibinfo {title}
  {{First-principles real-space linear-muffin-tin-orbital calculations of
  3\emph{d} impurities in Cu}},}\ }\href@noop {} {\bibfield  {journal}
  {\bibinfo  {journal} {Phys. Rev. B}\ }\textbf {\bibinfo {volume} {46}},\
  \bibinfo {pages} {14570--14577} (\bibinfo {year} {1992})}\BibitemShut
  {NoStop}%
\bibitem [{\citenamefont {Barbosa}\ \emph {et~al.}(2001)\citenamefont
  {Barbosa}, \citenamefont {Muniz}, \citenamefont {Costa},\ and\ \citenamefont
  {Mathon}}]{bmcm}%
  \BibitemOpen
  \bibfield  {author} {\bibinfo {author} {\bibfnamefont {L.}~\bibnamefont
  {Barbosa}}, \bibinfo {author} {\bibfnamefont {R.~B.}\ \bibnamefont {Muniz}},
  \bibinfo {author} {\bibfnamefont {A.~T.}\ \bibnamefont {Costa}}, \ and\
  \bibinfo {author} {\bibfnamefont {J}~\bibnamefont {Mathon}},\ }\bibfield
  {title} {\enquote {\bibinfo {title} {{Spin waves in ultrathin ferromagnetic
  overlayers}},}\ }\href@noop {} {\bibfield  {journal} {\bibinfo  {journal}
  {Phys. Rev. B}\ }\textbf {\bibinfo {volume} {63}},\ \bibinfo {pages} {174401}
  (\bibinfo {year} {2001})}\BibitemShut {NoStop}%
\bibitem [{\citenamefont {Kubo}(1957)}]{Kubo:1957cl}%
  \BibitemOpen
  \bibfield  {author} {\bibinfo {author} {\bibfnamefont {R.}~\bibnamefont
  {Kubo}},\ }\bibfield  {title} {\enquote {\bibinfo {title}
  {{Statistical-Mechanical Theory of Irreversible Processes. I. General Theory
  and Simple Applications to Magnetic and Conduction Problems}},}\ }\href@noop
  {} {\bibfield  {journal} {\bibinfo  {journal} {J. Phys. Soc. Jpn.}\ }\textbf
  {\bibinfo {volume} {12}},\ \bibinfo {pages} {570--586} (\bibinfo {year}
  {1957})}\BibitemShut {NoStop}%
\bibitem [{\citenamefont {Costa}\ \emph {et~al.}(2006)\citenamefont {Costa},
  \citenamefont {Muniz},\ and\ \citenamefont {Mills}}]{Costa:2006hb}%
  \BibitemOpen
  \bibfield  {author} {\bibinfo {author} {\bibfnamefont {A.~T.}\ \bibnamefont
  {Costa}}, \bibinfo {author} {\bibfnamefont {R.~B.}\ \bibnamefont {Muniz}}, \
  and\ \bibinfo {author} {\bibfnamefont {D.~L.}\ \bibnamefont {Mills}},\
  }\bibfield  {title} {\enquote {\bibinfo {title} {{Ferromagnetic resonance
  linewidths in ultrathin structures: A theoretical study of spin pumping}},}\
  }\href@noop {} {\bibfield  {journal} {\bibinfo  {journal} {Phys. Rev. B}\
  }\textbf {\bibinfo {volume} {73}},\ \bibinfo {pages} {054426} (\bibinfo
  {year} {2006})}\BibitemShut {NoStop}%
\bibitem [{\citenamefont {Costa}\ \emph {et~al.}(2010)\citenamefont {Costa},
  \citenamefont {Muniz}, \citenamefont {Lounis}, \citenamefont {Klautau},\ and\
  \citenamefont {Mills}}]{Costa2010}%
  \BibitemOpen
  \bibfield  {author} {\bibinfo {author} {\bibfnamefont {A.~T.}\ \bibnamefont
  {Costa}}, \bibinfo {author} {\bibfnamefont {R.~B.}\ \bibnamefont {Muniz}},
  \bibinfo {author} {\bibfnamefont {S.}~\bibnamefont {Lounis}}, \bibinfo
  {author} {\bibfnamefont {A.~B.}\ \bibnamefont {Klautau}}, \ and\ \bibinfo
  {author} {\bibfnamefont {D.~L.}\ \bibnamefont {Mills}},\ }\bibfield  {title}
  {\enquote {\bibinfo {title} {{Spin-orbit coupling and spin waves in ultrathin
  ferromagnets: The spin-wave Rashba effect}},}\ }\href@noop {} {\bibfield
  {journal} {\bibinfo  {journal} {Phys. Rev. B}\ }\textbf {\bibinfo {volume}
  {82}},\ \bibinfo {pages} {014428} (\bibinfo {year} {2010})}\BibitemShut
  {NoStop}%
\end{thebibliography}%

\section*{Acknowledgments}

We are very grateful to A. B. Klautau for providing the tight-binding parameters, to J. R. Suckert and S. Bl\"ugel for discussions, and for computing time on JURECA and JUQUEEN supercomputers at J\"ulich Supercomputing Centre. We thank the support CNPq (Brazil) and Alexander von Humboldt foundation (Germany). This work is supported by the HGF-YIG Programme VH-NG-717 (Functional Nanoscale Structure and Probe Simulation Laboratory-Funsilab) and the European Research Council (ERC) under the European Union's Horizon 2020 research and innovation programme (ERC-consolidator grant 681405 -- DYNASORE).

\section*{Author contributions}

F.S.M.G. performed calculations and wrote the manuscript. F.S.M.G., M.d.S.D., J.B., A.T.C., R.B.M. and S.L. analyzed and discussed the results. All authors reviewed the manuscript. 

\section*{Competing interests}

The authors declare no competing financial interests.


\clearpage

\begin{table}[ht]
\caption{Magnetoresistances $\Delta I^{\alpha,\beta,\gamma}=I'_{\omega,\|}(90^\circ)-I'_{\omega,\|}(0^\circ)$ for applied dc and resonant electric fields for the different rotation angles represented in the central panel of Fig.~\ref{fig:angular}.}
\begin{center}
\begin{tabular}{c|c|c|c|c|}
\cline{2-5}
& Frequency &\makecell{$\Delta I^{\alpha}$ \\ (xy plane)} & \makecell{$\Delta I^{\beta}$ \\ (zy plane)} & \makecell{$\Delta I^{\gamma}$ \\(zx plane) }\\ \cline{1-5}
\multirow{2}{*}{Co/Pt(001)}  & $\omega = 0$ (dc)                            & -1.86  & -0.99  & 0.89 \\ 
                                             & $\omega = \omega_\text{res}$ (ac) & -13.22  & -7.37 & 5.85 \\ \cline{1-5}
\multirow{2}{*}{Fe/W(110)}  & $\omega = 0$ (dc)                            & 0.52  & 0.91  & 0.39 \\ 
                                             & $\omega = \omega_\text{res}$ (ac) & -0.26   & 5.63  & 5.89 \\ \cline{1-5}
\end{tabular}
\end{center}
\label{tab:longitudinal}
\end{table}%

\begin{table}[ht]
\caption{Maximum amplitude of Hall currents for applied dc and resonant electric fields. 
When the magnetization is rotated in the $xy$ plane (i.e., PHE symmetry), we also indicate in parenthesis the angle for which the current is maximum. 
The minus sign represents the change from the reference case of static Co/Pt(001) currents.}
\begin{center}
\begin{tabular}{c|c|c|c|}
\cline{2-4}
& Frequency &$I'_{\omega,\perp}(\mathbf{m}_0 \in xy)$ & $I'_{\omega,\perp}(\mathbf{m}_0 \| \mathbf{\hat{z}})$\\ \cline{1-4}
\multirow{2}{*}{Co/Pt(001)}  & $\omega = 0$ (dc)                            & 0.05 ($\phi = 35^\circ$)  &  0.44 \\ 
                                             & $\omega = \omega_\text{res}$ (ac) & 0.22 ($\phi = 42^\circ$)  &  0.54 \\ \cline{1-4}
\multirow{2}{*}{Fe/W(110)}  & $\omega = 0$ (dc)                            & 0.09 ($\phi = 55^\circ$)  &  0.22 \\ 
                                             & $\omega = \omega_\text{res}$ (ac) & 0.36 ($\phi = 69^\circ$) & -0.29 \\ \cline{1-4}
\end{tabular}
\end{center}
\label{tab:transverse}
\end{table}%

\clearpage

\clearpage

\clearpage

\clearpage

\end{document}